\documentclass[twocolumn,trackchanges,dvipsnames]{aastex631}

\shorttitle{Self-Gravitating Hydrodynamics}
\shortauthors{Hanawa \& Mullen}

\graphicspath{{./}{figures/}}

\begin{document}

\title{Towards Higher Order Accuracy in Self-Gravitating Hydrodynamics}

\author[0000-0002-7538-581X]{Tomoyuki Hanawa}
\affiliation{Center for Frontier Science
1-33 Yayoi-cho, Inage-ku, Chiba, 263-8522, Japan}

\author[0000-0003-2131-4634]{Patrick D. Mullen}
\affiliation{CCS-2, Los Alamos National Laboratory, Los Alamos, NM 87545, USA}

\begin{abstract}
High order algorithms have emerged in numerical astrophysics as a promising avenue to reduce truncation error (proportional to a power of the linear resolution $\Delta x$) with only a moderate increase to computational expense.  Significant effort has been placed in the development of finite volume algorithms for (magneto)hydrodynamics, however, state-of-the-art astrophysical simulations tightly couple a plenitude of physics, additionally including gravity, photon transport, cosmic ray transport, chemistry, and/or diffusion, to name a few. 
Algorithms frequently operator split this additional physics (often a first order error in time) and/or adopt a model wherein their evaluation is limited to second order accuracy in space.  In this work, we present a fourth order accurate finite volume scheme for self-gravitating hydrodynamics on a uniform Cartesian grid. The method supplies source terms for the gravitational acceleration ($\rho \mathbf{g}$) and gravitational energy release ($\rho \mathbf{v} \cdot \mathbf{g}$) associated with fourth-order accurate solutions to the Poisson equation.  Our scheme (1) guarantees the conservation of total linear momentum, while (2) decreasing (in proportion to $\Delta x^4$) the effects of spurious heating and/or cooling associated with truncation error in the gravity.  We demonstrate expected convergence rates for the algorithm by measuring errors in test problems evolving self-gravity modified linear waves and 3D polytropic equilibria.  We test robustness of the algorithm by integrating an induced ``inside-out" adiabatic collapse.  We also discuss a method to smoothly downgrade the solution to second-order spatial accuracy to avoid spurious overshoots near steep density and/or pressure gradients.  
\end{abstract}


\section{Introduction} \label{sec:intro}

Numerical simulations have served as important tools to understand astronomical objects and phenomena for more than several decades, evolving large scale systems with complicated structures, subject to a diverse and complex set of physical processes (e.g., gravity, radiation, chemical reactions, etc.).  Often, these intricate simulations require many numerical cells to resolve fine scale features.  Higher order algorithms provide a pathway to accelerate numerical convergence for only a moderate increase in computational cost.  For this reason, such algorithms have been targeted for finite volume (magneto)hydrodynamics \citep[see, e.g.,][]{colella84, colella08, mccorquodale15, felker18}.

Self-gravity plays a fundamental role in determining the structure and evolution of various astronomical objects and hence has been taken into account in many numerical simulations \citep[e.g.,][to name a few]{couch13, katz16, kim18, mullen20, xu24}.  However, to our knowledge, all existing finite volume algorithms for self-gravitating hydrodynamics are limited to second order accuracy in space. Truncation error in the gravity may be serious when it is the main driver of the dynamics or if other parts of the simulation are all of higher order accuracy in space.  

Gravity appears in the (magneto)hydrodynamic equations as source terms, the gravitational acceleration and gravitational energy release, which are equated with changes in the momentum density and the total energy (excluding gravitational energy), respectively. The gravitational acceleration should conserve total linear momentum.  Numerically, conservation of total linear momentum can be guaranteed if the gravitational acceleration can be expressed as the divergence of a gravitational stress tensor \citep[see, e.g.,][]{shu88,jiang13,mullen21}. However, the divergence of the gravitational stress tensor might induce spurious circulation \citep[see][for more details]{mullen21}. An ideal, numerical evaluation of the gravitational acceleration should then (1) be accurate ($\mathcal{O} (\left[ \Delta x^r\right]$ with $r \geq 3$), (2) conserve total linear momentum, and (3) not promote anomalous forcing due to artificial circulation.  Similarly, the gravitational energy release should be of high order spatial accuracy, conserve total energy (including gravitational energy), and should avoid spurious heating and/or cooling due to truncation error in the gravity. The conservation of total energy does not guarantee the accuracy of the local thermal energy. Gravity changes the kinetic energy of a gas element and resulting gas motion converts some kinetic energy into thermal energy via compression or expansion, nevertheless, the specific entropy of the gas element should remain constant (in the absence of shocks). The truncation error in the specific entropy should be a fourth order small quantity if solutions using the high order gravitational acceleration and energy release are truly fourth order accurate in their discretized forms.

This paper is organized as follows.  We review the governing equations of self-gravitating hydrodynamics in \S2. We present our fourth order algorithm in \S3. In \S4, we apply the algorithm to several test problems including self-gravitating linear wave propagation, slab equilibrium advection, polytropic sphere equilibria evolution, and spherical gravitational collapse. In \S 5, we describe (1) robustness strategies about density and/or pressure discontinuities and (2) future potential improvements to our algorithm.  Finally, we summarize our results in \S 6.

This paper is based on earlier results presented in the 15th and 16th Conferences on Numerical Modeling of Space Plasma Flows \citep[see,][and Hanawa \& Mullen \textit{submitted}, for the 15th and 16th conferences, hereafter HM24 and HM25, respectively]{hanawa24}.  This paper includes technical details and additional examples omitted in the proceedings.

\section{Governing Equations}
The dynamical equations for self-gravitating hydrodyanmics are
\begin{eqnarray}
\frac{\partial \rho}{\partial t} + \mathbf{\nabla}\cdot\left[ \rho \mathbf{v} \right] & = & 0 , \\
\frac{\partial \rho \mathbf{v}}{\partial t} + \mathbf{\nabla} \cdot \left[ \rho \mathbf{vv} + P \mathbf{I} \right] & = & \rho \mathbf{g} , \label{momentum} \\
\frac{\partial E}{\partial t} + \mathbf{\nabla} \cdot \left [ (E + P) \mathbf{v} \right] & = & \rho \mathbf{v} \cdot \mathbf{g} , \label{energy}
\end{eqnarray}
where $\rho$ is the mass density, $\mathbf{v}$ is the velocity, $P$ is the gas pressure, $E = \rho \mathbf{v}^2 / 2 + e$ is the total energy (not including gravitational energy), and $e$ is the volumetric internal energy.  The gravity $\mathbf{g}$ is derived via the gradient of the gravitational potential $\phi$,
\begin{eqnarray}
\mathbf{g} & = & - \mathbf{\nabla} \phi,
\end{eqnarray}
and is therefore subject to the constraint $\nabla \times \mathbf{g} = 0$.  
The gravitational potential $\phi$ is related to the mass density $\rho$ via the Poisson equation
\begin{eqnarray}
\mathbf{\nabla}^2 \phi & = & 4 \pi G \rho,  \label{poisson}
\end{eqnarray}
where $ G$ denotes the gravitational constant.

The specific entropy $s$ is constant along streamlines
\begin{eqnarray}
\frac{\partial s}{\partial t} + \mathbf{v}\cdot \mathbf{\nabla s} & \ge & 0,
\label{eq::spec_entr}
\end{eqnarray}
when neglecting any interplay with magnetic fields and/or heating/cooling sources (HM25).  Albeit our proposed algorithm is straightforwardly extensible to general equations of state (EOS), we assume ideality in this work, hence permitting us to relate
\begin{eqnarray}
e & = & \frac{P}{\gamma - 1} , \\
s & = & \ln P - \gamma \ln \rho,
\end{eqnarray}
where $ \gamma $ denotes the adiabatic index.  

Building on previous work \cite[see, e.g.,][]{shu91,jiang13,mullen21}, the momentum equation can be rewritten as 
\begin{equation}
\frac{\partial \rho \mathbf{v}}{\partial t}  + \mathbf{\nabla} \cdot \left[ \rho \mathbf{vv} + P \mathbf{I} + \mathbf{T} _{g} \right]  =  0 , 
\label{eq:hydro_with_tg}
\end{equation}
where 
\begin{eqnarray}
\mathbf{T} _{g} & = & \frac{1}{4 \pi G}
\left( \mathbf{gg} - \frac{\mathbf{g} ^2}{2} \mathbf{I} \right) , \label{momentum2}
\end{eqnarray}
is the gravitational stress tensor.  Equation (\ref{eq:hydro_with_tg}) highlights that the gravity supplemented momentum equation can still be formulated as a conservation law \textit{without} a source term, thereby demonstrating conservation of linear momentum. The divergence of the gravitational stress tensor is
\begin{eqnarray}
- \nabla \cdot \mathbf{T_g} = -\frac{1}{4 \pi G} \left[(\nabla \cdot \mathbf{g}) \mathbf{g} + (\nabla \times \mathbf{g}) \times \mathbf{g} \right],
\label{eq::divtg_exp}
\end{eqnarray}
and equivalence of Equations (\ref{momentum}) and (\ref{eq:hydro_with_tg}) implies
\begin{eqnarray}
\rho \mathbf{g} & = & - \mathbf{\nabla} \cdot \mathbf{T} _g  \label{divTg},
\end{eqnarray}
under the condition $\nabla \times \mathbf{g} = 0$.  

The total energy equation can similarly be recast \citep[see, e.g.,][]{hanawa19,mullen21} as
\begin{equation}
\frac{\partial E_\mathrm{tot}}{\partial t}
+ \mathbf{\nabla} \cdot \left[(E + P) \mathbf{v} + \mathbf{F} _{g} \right]  =  0 ,
\label{energy2}
\end{equation}
where
\begin{eqnarray}
\mathbf{F} _{g} & = & \phi \left( \rho \mathbf{v} - \frac{1}{4\pi G}\frac{\partial \mathbf{g}}{\partial t} \right) 
\end{eqnarray}
is the gravitational energy flux and
\begin{eqnarray}
E_\mathrm{tot} = E + E_g
\end{eqnarray}
is the total energy \textit{including} the gravitational energy
\begin{eqnarray}
E_g = - \frac{\mathbf{g} ^2}{8 \pi G}.
\label{Egrav}
\end{eqnarray}

\section{Discretization}

Next, we discretize the governing equations given in the previous section.  We assume a uniform Cartesian grid with linear resolution $ h $ in the $x$-, $y$-, and $z$- directions, where each cell center is assigned a spatial position  
\begin{equation}
\left( x _i, y _j, z _k \right) = h \left(i + \frac{1}{2},  j + \frac{1}{2}, k + \frac{1}{2} \right) ,
\end{equation}
where $ i $, $ j $, and $ k $ cell indices map logical space to Cartesian $x$, $y$ and $z$ space, respectively. 

Following HM25, the cell-volume-averaged gravitational potential and mass density are
\begin{eqnarray}
\phi _{i,j,k} & = & \frac{1}{h^3} \int _{z_k-\frac{h}{2}} ^{z_k+\frac{h}{2}} \int _{y_j-\frac{h}{2}} ^{y_j+\frac{h}{2}} \int _{x_i-\frac{h}{2}} ^{x _i+\frac{h}{2}} \phi (x, y, z) \nonumber \\ & & dx dy dz , \label{phi_ijk_def} \\
\rho _{i,j,k} & = & \frac{1}{h^3} \int _{z_k-\frac{h}{2}} ^{z_k+\frac{h}{2}} \int _{y_j-\frac{h}{2}} ^{y_j+\frac{h}{2}} \int _{x_i-\frac{h}{2}} ^{x _i+\frac{h}{2}} \rho (x, y, z) \nonumber \\
& &dx dy dz,
\end{eqnarray}
and the cell-surface-averaged interface gravity is
\begin{eqnarray}
\left\langle g _x \right\rangle _{i+\frac{1}{2},j,k} & = & - \frac{1}{h^2} 
\int _{z_k-\frac{h}{2}} ^{z _k+\frac{h}{2}} \int _{y_j-\frac{h}{2}} ^{y_j+\frac{h}{2}} \frac{\partial \phi}{\partial x} \left(x _{i+\frac{1}{2}}, y, z \right) \nonumber \\ && dy dz,
\end{eqnarray}
where we have used a half integer to designate quantities evaluated at a cell face (i.e., $x _{i+\frac{1}{2}} = x _i + \frac{h}{2}$) and angle decorators $\left< \cdot \right>$ to denote an average over the cell surface.  

Applying a central difference of fourth order accuracy in all directions, we obtain the discretized Poisson equation
\begin{eqnarray}
\frac{- \phi _{i+2,j,k}  +  16 \phi _{i+1,j,k} - 30 \phi _{i,j,k} + 16 \phi _{i-1,j,k} - \phi _{i-2,j,k}}{12} \nonumber \\
+ \frac{- \phi _{i,j+2,k}  +  16 \phi _{i,j+1,k} - 30 \phi _{i,j,k} + 16 \phi _{i,j-1,k} - \phi _{i,j-2,k}}{12} \nonumber \\
+ \frac{- \phi _{i,j,k+2}  + 16 \phi _{i,j,k+1} - 30 \phi _{i,j,k} + 16 \phi _{i,j,k-1} - \phi _{i,j,k-2} }{12} \nonumber \\
 =  4 \pi G h ^2 \rho _{i,j,k} . \hskip 120pt \label{dPoisson}
\end{eqnarray}
Solutions to Equation (\ref{dPoisson}) can be obtained via fast Fourier transforms (hereafter FFTs) or multigrid \citep[see, e.g.,][]{trottenberg00, tomida23}.

We construct high order gravitational accelerations and gravitational energy releases via linear combinations of second order accurate gravitational accelerations
\begin{eqnarray}
g _{x,i+\frac{1}{2},j,k} & = & - \frac{\phi _{i+1,j,k}-\phi_{i,j,k}}{h} \label{gx} \\
& = & \langle g _x \rangle _{i+\frac{1}{2},j,k} + {\cal O} \left(h ^2 \right), \\
g _{y,i,j+\frac{1}{2},k} & = & - \frac{\phi _{i,j+1,k}-\phi_{i,j,k}}{h}  \label{gy} \\
& = & \langle g _y \rangle _{i,j+\frac{1}{2},k} + {\cal O} \left(h ^2 \right),  \\
g _{z,i,j,k+\frac{1}{2}} & = & - \frac{\phi _{i,j,k+1}-\phi_{i,j,k}}{h} \label{gz} \\
& = & \langle g _z \rangle _{i,j,k+\frac{1}{2}} + {\cal O} \left(h ^2 \right).
\end{eqnarray}
The fourth order accurate gravity evaluated on the cell surface centered at $ \left(x _{i+\frac{1}{2}}, y _j, z _k \right)$ is
\begin{eqnarray}
g _{x,i+\frac{1}{2},j,k} ^{(4)} & \equiv & 
\frac{- g _{x,i+\frac{3}{2},j,k} + 14 g _{x,i+\frac{1}{2},j,k} - g _{x,i-\frac{1}{2},j,k}}{12} \\
& = & \left\langle g _x \right\rangle _{i+\frac{1}{2},j,k} + {\cal O} \left( h ^4 \right).
\end{eqnarray}
The $y$- and $z$- components of the gravity (on their respective cell surfaces) are
\begin{eqnarray}
g _{y,i,j+\frac{1}{2},k} ^{(4)} & \equiv & 
\frac{- g _{y,i,j+\frac{3}{2},k} + 14 g _{y,i,j+\frac{1}{2},k} - g _{y,i,j-\frac{1}{2},k}}{12} , \\
g _{z,i,j,k+\frac{1}{2}} ^{(4)} & \equiv & 
\frac{- g _{z,i,j,k+\frac{3}{2}} + 14 g _{z,i,j,k+\frac{1}{2}} - g _{z,i,j,k-\frac{1}{2}}}{12} .
\end{eqnarray}
We can then rewrite Equation (\ref{dPoisson}) as
\begin{eqnarray}
-\frac{g _{x,i+\frac{1}{2},j,k} ^{(4)} -g _{x,i-\frac{1}{2},j,k} ^{(4)}}{h} & - & \frac{g _{y,i,j+\frac{1}{2},k} ^{(4)} -g _{y,i,j-\frac{1}{2},k} ^{(4)}}{h} \nonumber \\ 
- \frac{g _{z,i,j,k+\frac{1}{2}} ^{(4)} -g _{z,i,j,k-\frac{1}{2}} ^{(4)}}{h} & = & 4 \pi G \rho _{i,j,k}. \label{dPoisson2}
\end{eqnarray}
One could draw parallels between the left hand side of Equation (\ref{dPoisson2}) and the typical numerical application of Gauss' law in a finite volume construction, wherein $g _{x,i+\frac{1}{2},j,k} ^{(4)}$ is analogous to the numerical mass flux $\left< \rho \mathbf{v} \right>_{x,i+1/2,j,k}$ in the continuity equation.

\begin{widetext}
We emphasize that $ \displaystyle \frac{1}{h^3} \int_V \left( \rho \mathbf{g} \right) dV \neq \rho^{(4)} \mathbf{g}^{(4)} + \mathcal{O}\left(h^4 \right)$.  To proceed, we derive a high order gravitational acceleration ($\rho \mathbf{g}$) based on Equation (\ref{divTg}). \cite{mullen21} presented the second order accurate gravitational stress tensor
\begin{eqnarray}
T _{g,xx,i+\frac{1}{2},j,k} ^{(2)} & = & \frac{1}{8\pi G} g _{x,i+\frac{1}{2},j,k} ^2 
- \frac{1}{16\pi G} g _{y,i,j+\frac{1}{2},k} g _{y,i+1,j+\frac{1}{2},k} 
- \frac{1}{16\pi G}  g _{y,i,j-\frac{1}{2},k} g _{y,i+1,j-\frac{1}{2},k}  \nonumber \\
&& \hspace{62pt} - \frac{1}{16\pi G} g _{z,i,j,k+\frac{1}{2}} g _{z,i+1,j,k+\frac{1}{2}} 
- \frac{1}{16\pi G}  g _{z,i,j,k-\frac{1}{2}} g _{z,i+1,j,k-\frac{1}{2}}  , \label{txx2} \\
T _{g, yx,i,j+\frac{1}{2},k}^{(2)}  & = & \frac{g _{y,i,j+\frac{1}{2},k}}{16 \pi G} \left( g _{x,i+\frac{1}{2},j,k} + g _{x,i+\frac{1}{2},j+1,k} + g _{x,i-\frac{1}{2},j,k} + g _{x,i-\frac{1}{2},j+1,k}\right) , \label{tyx2}\\
T _{g, zx,i,j,k+\frac{1}{2}}^{(2)}  & = & \frac{g _{z,i,j,k+\frac{1}{2}}}{16 \pi G} \left( g _{x,i+\frac{1}{2},j,k} + g _{x,i+\frac{1}{2},j,k+1} + g _{x,i-\frac{1}{2},j,k} + g _{x,i-\frac{1}{2},j,k+1}\right) , \label{tzx2}
\end{eqnarray}
which provides the second order accurate gravitational acceleration
\begin{eqnarray}
\left( \rho g _x \right) _{i,j,k} ^{(2)} & = & -
\frac{T _{g,xx,i+\frac{1}{2},j,k} ^{(2)} -T_{g,xx,i-\frac{1}{2},j,k} ^{(2)}}{h} 
- \frac{T _{g,yx,i,j+\frac{1}{2},k} ^{(2)}-T_{g,yx,i,j-\frac{1}{2},k} ^{(2)}}{h}
- \frac{T _{g,zx,i,j,k+\frac{1}{2}} ^{(2)}-T_{g,zx,i,j,k-\frac{1}{2}} ^{(2)}}{h} \\
& = & \frac{1}{2} \rho _{i,j,k} ^{(2)} \left( g _{x,i+\frac{1}{2},j,k} + g _{x,i-\frac{1}{2},j,k}\right) , \label{rhogx2} \\
\rho _{i,j,k} ^{(2)} & = & \frac{-g_{x,i+\frac{1}{2},j,k} + g _{x,i-\frac{1}{2},j,k}}{4\pi G h} 
+ \frac{-g_{y,i,j+\frac{1}{2},k} + g _{y,i,j-\frac{1}{2},k}}{4\pi G h} 
+ \frac{-g_{z,i,j,k+\frac{1}{2}} + g _{z,i,j,k-\frac{1}{2}}}{4\pi G h} .
\end{eqnarray}
Remaining components of the gravitational stress tensor and gravitational acceleration can be obtained by replacing coordinates and indices cyclically.

We now refine the gravitational stress tensor into a fourth order accurate one. In the following, we restrict ourselves to the cell surface in the $x$-direction to save unnecessary repetition. First we evaluate the cell-surface-averaged gravitational stress tensor $ \langle T _{xx} \rangle _{i+\frac{1}{2},j,k} $  via Taylor series expansion about the cell surface center,
\begin{eqnarray}
\left\langle T _{g,xx} \right\rangle _{i+\frac{1}{2},j,k} 
& = & \frac{1}{8\pi G} \left\langle g _x {}^2 - g _y {}^2 - g _z {}^2 \right\rangle _{i+\frac{1}{2},j,k} , \label{txx-n} \\
\langle g _x ^2 \rangle _{i+\frac{1}{2},j,k} & \equiv &
\frac{1}{h^2} \int _{z_k-\frac{h}{2}} ^{z_k+\frac{h}{2}} \int _{y_j-\frac{h}{2}} ^{y_j+\frac{h}{2}} 
\left[ \frac{\partial \phi}{\partial x} \left( x _{i+\frac{h}{2},} y, z \right) \right] ^2 dy dz \\
& = &  \left( \frac{\partial \phi}{\partial x} \right) ^2 + \frac{h^2}{12} \left[ \frac{\partial \phi}{\partial x} \left( \frac{\partial ^3 \phi}{\partial x \partial y^2} + \frac{\partial ^3 \phi}{\partial x \partial z^2} \right) + \left( \frac{\partial ^2\phi}{\partial x \partial y} \right) ^2 + \left( \frac{\partial ^2\phi}{\partial x \partial z} \right) ^2 \right] + {\cal O} \left(h^4 \right), \\
\left\langle g _y ^2 \right\rangle _{i+\frac{1}{2},j,k} & = &
\frac{1}{h^2} \int _{z_k-\frac{h}{2}} ^{z_k+\frac{h}{2}} \int _{y_j-\frac{h}{2}} ^{y_j+\frac{h}{2}} 
\left[ \frac{\partial \phi}{\partial y} \left( x _{i+\frac{h}{2},} y, z \right) \right] ^2 dy dz \\
& = & \left( \frac{\partial \phi}{\partial y} \right) ^2 + \frac{h^2}{12} \left[ \left( \frac{\partial^2 \phi}{\partial y^2} \right) ^2 + \left( \frac{\partial ^2 \phi}{\partial y \partial z} \right)^2 + \frac{\partial \phi}{\partial y} \left( \frac{\partial^3 \phi}{\partial y^3} + \frac{\partial ^3 \phi}{\partial y \partial z^2} \right)\right] + {\cal O} \left(h ^4\right), \\
\left\langle g _z ^2 \right\rangle _{i+\frac{1}{2},j,k}  & = &
\frac{1}{h^2} \int _{z_k-\frac{h}{2}} ^{z_k+\frac{h}{2}} \int _{y_j-\frac{h}{2}} ^{y_j+\frac{h}{2}} 
\left[ \frac{\partial \phi}{\partial z} \left( x _{i+\frac{h}{2},} y, z \right) \right] ^2 dy dz \\
& = & \left( \frac{\partial \phi}{\partial z} \right) ^2 + \frac{h^2}{12} \left[ \left( \frac{\partial^2 \phi}{\partial z^2} \right) ^2 + \left( \frac{\partial ^2 \phi}{\partial y \partial z} \right)^2  + \frac{\partial \phi}{\partial z} \left( \frac{\partial^3 \phi}{\partial z^3} +  \frac{\partial ^3 \phi}{\partial z \partial y^2}\right)\right] + {\cal O} \left( h ^4 \right) . \label{gzz-n}
\end{eqnarray}
We remark that HM24 Equations (27)--(29),  which should be equivalent to Equations (\ref{txx-n})--(\ref{gzz-n}), contain several typos.
Next, we similarly evaluate the off-diagonal components of the gravitational
tensor,
\begin{eqnarray}
\left\langle T _{g,xy} \right\rangle _{i+\frac{1}{2},j,k} 
& = & \frac{1}{4\pi G} \left\langle g _x g _y \right\rangle _{i+\frac{1}{2},j,k} , \label{txy-n} \\
\left\langle T _{g,xz} \right\rangle _{i+\frac{1}{2},j,k} 
& = & \frac{1}{4\pi G} \left\langle g _x g _z \right\rangle _{i+\frac{1}{2},j,k} , \label{txz-n} \\
\left\langle g _x g _y \right\rangle _{i+\frac{1}{2},j,k} & = &
\frac{1}{h^2} \int _{z_k-\frac{h}{2}} ^{z_k+\frac{h}{2}} \int _{y_j-\frac{h}{2}} ^{y_j+\frac{h}{2}}
\frac{\partial \phi}{\partial x} \left( x _{i+\frac{h}{2},} y, z \right) \frac{\partial \phi}{\partial y} \left( x _{i+\frac{h}{2},} y, z \right) dy dz \\
& = & \frac{\partial \phi}{\partial x} \frac{\partial \phi}{\partial y} + \frac{h^2}{12} 
\left[ \frac{\partial ^2 \phi}{\partial x \partial y} \frac{\partial ^2 \phi}{\partial y^2} + \frac{\partial ^2 \phi}{\partial x \partial z} \frac{\partial ^2 \phi}{\partial y \partial z} + \frac{1}{2}\frac{\partial \phi}{\partial x} \left(\frac{\partial^3 \phi}{\partial y ^3 }+ \frac{\partial ^3 \phi}{\partial y \partial z ^2} \right) \right. \nonumber \\ 
& & \hspace{60pt} \left. 
+ \frac{1}{2}\frac{\partial \phi}{\partial y} \left(\frac{\partial^3 \phi}{\partial x\partial y ^2}+ \frac{\partial ^3 \phi}{\partial x \partial z ^2} \right) \right] + {\cal O} \left( h ^4 \right) , \\
\left\langle g _x g _z \right\rangle _{i+\frac{1}{2},j,k} & = &
\frac{1}{h^2}  \int _{z_k-\frac{h}{2}} ^{z_k+\frac{h}{2}} \int _{y_j-\frac{h}{2}} ^{y_j+\frac{h}{2}} 
\frac{\partial \phi}{\partial x} \left( x _{i+\frac{h}{2},} y, z \right) \frac{\partial \phi}{\partial z} \left( x _{i+\frac{h}{2},} y, z \right) dy dz \\
& = & \frac{\partial \phi}{\partial x} \frac{\partial \phi}{\partial z} + \frac{h^2}{12} 
\left[ \frac{\partial ^2 \phi}{\partial x \partial z} \frac{\partial ^2 \phi}{\partial z^2} + \frac{\partial ^2 \phi}{\partial x \partial y} \frac{\partial ^2 \phi}{\partial y \partial z} + \frac{1}{2}\frac{\partial \phi}{\partial x} \left(\frac{\partial^3 \phi}{\partial y ^3 }+ \frac{\partial ^3 \phi}{\partial y \partial z ^2} \right)\right. \nonumber \\
&& \hspace{60pt}\left. + \frac{1}{2}\frac{\partial \phi}{\partial y} \left(\frac{\partial^3 \phi}{\partial x\partial y ^2}+ \frac{\partial ^3 \phi}{\partial x \partial z ^2} \right) \right] + {\cal O} \left( h ^4 \right) ,
\end{eqnarray}

Next, we obtain the difference between $ \langle T _{g, xx} \rangle _{i+\frac{1}{2},j,k} $ and $ T _{g, xx,i+\frac{1}{2},j,k} ^{(2)} $; these correspond to \lq\lq fourth order correction terms." To arrive at these correction terms, we (i) recast the cell-averaged gravitational potential via Taylor series expansion around $ \left(x _{i+\frac{1}{2}}, y_j, z _k \right) $,
\begin{eqnarray}
\phi (x, y, z) & = & \phi \left( x _{i+\frac{1}{2}}, y _j, z _k \right) + \frac{\partial \phi}{\partial x} \left( x - x _{i+\frac{1}{2}} \right) + \frac{\partial \phi}{\partial y}
\left( y - y _j \right) + \frac{\partial \phi}{\partial z} \left( z - z _k \right)  \nonumber \\
& &+ \frac{1}{2} \frac{\partial ^2 \phi}{\partial x ^2} \left( x - x _{i+\frac{1}{2}} \right) ^2 + \frac{1}{2} \frac{\partial ^2 \phi}{\partial y ^2} \left( y - y _j \right) ^2
+ \frac{1}{2} \frac{\partial \phi ^2}{\partial z ^2} \left( z - z _k \right) ^2 \nonumber \\
& & + \frac{\partial ^2 \phi}{\partial x \partial y} \left(x - x _{i+\frac{1}{2}} \right) \left( y - y _j \right) + \frac{\partial ^2 \phi}{\partial y} \left(y - y _j \right) \left( z - z _k \right) + 
\frac{\partial ^2 \phi}{\partial z \partial x} \left( z - z _k\right) \left( x - x _{i+\frac{1}{2}} \right) \nonumber \\
&& + \frac{1}{6}\frac{\partial ^3 \phi}{\partial x ^3} \left( x - x _{i+\frac{1}{2}} \right) ^3
+ \frac{1}{6}\frac{\partial ^3 \phi}{\partial y ^3} \left( y - y _j\right) ^3
+ \frac{1}{6} \frac{\partial ^3 \phi}{\partial z ^3} \left( z - z _k\right) ^3 \nonumber \\
& &+ \frac{1}{2}\frac{\partial ^3 \phi}{\partial x ^2 \partial y} \left(x - x _{i+\frac{1}{2}} \right) ^2 \left( y - y _j \right) + \frac{1}{2}\frac{\partial ^3 \phi}{\partial y ^2 \partial z} \left(y - y _j \right) ^2 \left( z - z _k \right) \nonumber \\
& & + \frac{1}{2}\frac{\partial ^3 \phi}{\partial z ^2 \partial x} \left(z - z _k \right) ^2 \left( x - x _{i+\frac{1}{2}} \right) + \frac{\partial ^3 \phi}{\partial x \partial y \partial z} \left(x - x _{i+\frac{1}{2}}\right) \left( y - y _j \right) \left( z - z _k \right) .  \label{Taylor}
\end{eqnarray}
Substituting Equation (\ref{Taylor}) into Equation (\ref{phi_ijk_def}) we obtain
\begin{eqnarray}
\phi _{i,j,k} & = & \phi (x _{i+\frac{1}{2}}, y _j,z _k) - \frac{h}{2} \frac{\partial \phi}{\partial x} + \frac{h^2}{6} \frac{\partial^2 \phi}{\partial x ^2} + \frac{h^2}{24}  \frac{\partial ^2 \phi}{\partial y^2} + \frac{h^2}{24} \frac{\partial ^2 \phi}{\partial z^2} - \frac{h^3}{24} \frac{\partial ^3\phi}{\partial x^3} - \frac{h^3}{48} \frac{\partial ^3 \phi}{\partial x \partial y ^2}  - \frac{h^3}{48} \frac{\partial ^3 \phi}{\partial x \partial z ^2}  + {\cal O} (h^4) , \label{phi_ijk} \\ 
\phi _{i+1,j,k} & = & \phi (x _{i+\frac{1}{2}}, y _j,z _k) + \frac{h}{2} \frac{\partial \phi}{\partial x} + \frac{h^2}{6} \frac{\partial^2 \phi}{\partial x ^2} + \frac{h^2}{24}  \frac{\partial ^2 \phi}{\partial y^2} + \frac{h^2}{24} \frac{\partial ^2 \phi}{\partial z^2} + \frac{h^3}{24} \frac{\partial ^3\phi}{\partial x^3}  - \frac{h^3}{48} \frac{\partial ^3 \phi}{\partial x \partial y ^2}  - \frac{h^3}{48} \frac{\partial ^3 \phi}{\partial x \partial z ^2}  + {\cal O} (h^4) , \label{phi_i+1jk} \\
\phi _{i,j\pm 1,k} & = & \phi (x _{i+\frac{1}{2}}, y _j,z _k) - \frac{h}{2} \frac{\partial \phi}{\partial x} \pm h \frac{\partial \phi}{\partial y} + \frac{h^2}{6} \frac{\partial^2 \phi}{\partial x ^2} \mp \frac{h^2}{2} \frac{\partial ^2 \phi}{\partial x \partial y} + \frac{13 h^2}{24}  \frac{\partial ^2 \phi}{\partial y^2} + \frac{h^2}{24} \frac{\partial ^2 \phi}{\partial z^2} - \frac{h^3}{24} \frac{\partial ^3\phi}{\partial x^3} \pm \frac{h^3}{6} \frac{\partial ^3\phi}{\partial x^2 \partial y} \nonumber \\
&& \hspace{65pt} - \frac{h^3}{48} \frac{\partial ^3 \phi}{\partial x \partial y ^2} \pm \frac{5 h^3}{24} \frac{\partial ^3 \phi}{\partial y ^3} - \frac{h^3}{48} \frac{\partial ^3 \phi}{\partial x \partial z ^2} \pm \frac{h^3}{24} \frac{\partial ^3 \phi}{\partial y \partial z^2} + {\cal O} (h^4) , \label{phi_ijpm1k} \\
\phi _{i,j,k\pm 1} & = & \phi (x _{i+\frac{1}{2}}, y _j,z _k) - \frac{h}{2} \frac{\partial \phi}{\partial x} \pm h \frac{\partial \phi}{\partial z} + \frac{h^2}{6} \frac{\partial^2 \phi}{\partial x ^2} \mp \frac{h^2}{2} \frac{\partial ^2 \phi}{\partial x \partial z} + \frac{13 h^2}{24}  \frac{\partial ^2 \phi}{\partial z^2} + \frac{h^2}{24} \frac{\partial ^2 \phi}{\partial y^2} - \frac{h^3}{24} \frac{\partial ^3\phi}{\partial x^3} \pm \frac{h^3}{6} \frac{\partial ^3\phi}{\partial x^2 \partial z} \nonumber \\
&& \hspace{65pt} - \frac{h^3}{48} \frac{\partial ^3 \phi}{\partial x \partial z ^2} \pm \frac{5 h^3}{24} \frac{\partial ^3 \phi}{\partial z ^3} - \frac{h^3}{48} \frac{\partial ^3 \phi}{\partial x \partial y ^2} \pm \frac{h^3}{24} \frac{\partial ^3 \phi}{\partial z \partial y^2} + {\cal O} (h^4) , \label{phi_ijkpm1} 
\end{eqnarray}
then (ii) recast the second order gravity in Equations (\ref{gx})--(\ref{gz}) by substituting Equations (\ref{phi_ijk}) through (\ref{phi_ijkpm1}). 
\begin{eqnarray}
g _{x,i+\frac{1}{2},j,k} & = & 
 - \frac{\partial \phi}{\partial x}
- \frac{h^2}{12} \frac{\partial ^3 \phi}{\partial x^3}
- \frac{h^2}{24} \frac{\partial ^3 \phi}{\partial x \partial y ^2} 
- \frac{h^2}{24} \frac{\partial ^3 \phi}{\partial x \partial z ^2} + {\cal O} \left( h ^4 \right) , \label{gxT2} \\
g _{y,i,j\pm\frac{1}{2},k} & = & - \frac{\partial \phi}{\partial y} 
+ \frac{h}{2} \frac{\partial ^2 \phi}{\partial x \partial y} 
\mp \frac{h}{2} \frac{\partial ^2 \phi}{\partial y ^2} 
- \frac{h^2}{6} \frac{\partial ^3 \phi}{\partial x ^2 \partial y} 
\pm \frac{h^2}{4}\frac{\partial ^3 \phi}{\partial x \partial y ^2}  - \frac{5 h^2}{24} \frac{\partial ^3 \phi}{\partial y ^3}
- \frac{h^2}{24} \frac{\partial ^3 \phi}{\partial y \partial z ^2} + {\cal O} \left( h^3 \right), \label{gyT2} \\
g _{z,i,j,k\pm\frac{1}{2}} & = & - \frac{\partial \phi}{\partial z} 
+ \frac{h}{2} \frac{\partial ^2 \phi}{\partial y \partial z} 
\mp \frac{h}{2} \frac{\partial ^2 \phi}{\partial z ^2} 
- \frac{h^2}{6} \frac{\partial ^3 \phi}{\partial x ^2 \partial z} 
\pm \frac{h^2}{4}\frac{\partial ^3 \phi}{\partial x \partial z ^2} - \frac{5 h^2}{24} \frac{\partial ^3 \phi}{\partial z ^3}
- \frac{h^2}{24} \frac{\partial ^3 \phi}{\partial y ^2 \partial z} + {\cal O} \left( h^3 \right), \label{gzT2}
\end{eqnarray}
and finally (iii) substitute Equations (\ref{gxT2})--(\ref{gzT2}) into Equations (\ref{txx2})--(\ref{tzx2}) to express $ T _{g,xx,i+\frac{1}{2},j,k} ^{(2)} $ as a Taylor series expansion.  Note that all derivatives denote the values at  $ \left(x _{i+\frac{1}{2}}, y_j, z _k \right) $ (see Equation \ref{Taylor}).  Appendix A presents the difference between $ \langle T _{g,xx} \rangle _{i+\frac{1}{2},j,k} $ and $ T _{g,xx,i+\frac{1}{2},j,k} ^{(2)} $  up to fourth order small quantities. 
All correction terms are proportional to the square of the cell width ($\propto h ^2$). Since both the second and fourth order accurate gravitational stress tensors are symmetric with respect to inversion in the $ x $-, $ y $- and $ z $-directions, the difference does not include terms $ \propto h ^3 $ and the leading truncation error is $ \propto h ^4$.

As shown in Appendix A, we can express the correction terms by quadratics of the discretized gravity, $ g _{x,i+\frac{1}{2},j,k} $, $ g _{y,i,j+\frac{1}{2},k} $, and $ g _{z,i,j,k+\frac{1}{2}} $, however, each correction term can be expressed in several ways (see, e.g., HM24 for a different expression).  Similarly, \cite{mullen21} presented two different forms of the gravitational stress tensor, both of which were second order accurate, however, one of them induced spurious circulation in the gravitational acceleration in an envelope surrounding a spherical massive body, while the other did not. \cite{mullen21} found that the anomalous accelerations associated with the problematic discretization of the gravitational stress tensor were due to a non-zero rotation of the gravity.  Therefore, in short, we inform choices between various discretizations of the fourth order accurate correction terms by considering selections where $ \oint \mathbf{g} \cdot d\mathbf{s} = 0 $ vanishes along circuits such as,
\begin{eqnarray}
(x _i, y _j, z _k) \rightarrow (x _{i+1}, y _j, z _k) \rightarrow
(x _{i+1}, y _{j+1}, z _k) \\
\rightarrow (x _i, y _{j+1}, z _k) \rightarrow (x _{i-1}, y _{j+1}, z _k) 
\rightarrow (x _{i-1}, y _{j}, z _k) \rightarrow (x _i, y _j, z _k) .  
\end{eqnarray}
We refer interested readers to Appendix B where we supplement this discussion with more technical details and caveats surrounding $ \mathbf{\nabla} \times \mathbf{g} $ considerations in the discretized correction terms. 

Finally, we take the divergence of the fourth order accurate gravitational stress tensor to arrive at the high order accurate gravitational acceleration
\begin{eqnarray}
\left( \rho g _x \right) _{i,j,k} & = & \frac{\rho _{i,j,k}}{12} \left( - g _{x,i+\frac{3}{2},j,k} + 7 g _{x,i+\frac{1}{2},j,k} + 7 g _{x,i-\frac{1}{2},j,k} - g _{x,i-\frac{3}{2},j,k} \right) \nonumber \\
& & + \frac{\left( \rho _{i+1,j,k} - \rho _{i-1,j,k}\right)}{24} \left( g _{x,i+\frac{1}{2},j,k} - g _{x,i-\frac{1}{2},j,k} \right) \nonumber \\
& & + \frac{(\rho _{i,j+1,k} - \rho _{i,j,k})}{48} \left( g _{x,i+\frac{1}{2},j+1,k} + g _{x,i-\frac{1}{2},j+1,k} - g _{x,i+\frac{1}{2},j,k} - g _{x,i-\frac{1}{2},j,k} \right) \nonumber \\
& & + \frac{(\rho _{i,j,k} - \rho _{i,j-1,k})}{48} \left( g _{x,i+\frac{1}{2},j,k} + g _{x,i-\frac{1}{2},j,k} - g _{x,i+\frac{1}{2},j-1,k} - g _{x,i-\frac{1}{2},j-1,k} \right) \nonumber \\
& & + \frac{(\rho _{i,j,k+1} - \rho _{i,j,k})}{48} \left( g _{x,i+\frac{1}{2},j,k+1} + g _{x,i-\frac{1}{2},j,k+1} - g _{x,i+\frac{1}{2},j,k} - g _{x,i-\frac{1}{2},j,k} \right) \nonumber \\
& & + \frac{(\rho _{i,j,k} - \rho _{i,j,k-1})}{48} \left( g _{x,i+\frac{1}{2},j,k} + g _{x,i-\frac{1}{2},j,k} - g _{x,i+\frac{1}{2},j,k-1} - g _{x,i-\frac{1}{2},j,k-1} \right).
\label{rhogx4b}
\end{eqnarray}
\end{widetext}
Equation (\ref{rhogx4b}) contains contributions from products of (i) the density and gravity and (ii) their gradients.

In the presentation of Equation (\ref{rhogx4b}), we have taken the liberty of replacing occurrences of $\rho_{i,j,k}^{(2)}$ with $\rho_{i,j,k}$ (where the former is the quantity truly appearing following our derivation from the gravitational stress tensor).  This nicety saves the computational expense of computing $\rho_{i,j,k}^{(2)}$, while also being directly applicable to Poisson solutions invoking Jeans swindle,
\begin{eqnarray}
\nabla^2 \phi & = & 4 \pi G \left( \rho - \bar{\rho} \right),   \label{poissonJ} 
\end{eqnarray}
where $ \bar{\rho} $ denotes a mean density. The cell-volume-averaged mass density $\rho_{i,j,k}$ can be used in place of $\rho_{i,j,k}^{(2)}$ because the accuracy remains unchanged between the two prescriptions since they only differ in fourth order small quantities.  Linear momentum conservation is guaranteed when the source term invokes $\rho_{i,j,k}^{(2)}$ since the acceleration was derived from a gravitational stress tensor.  As shown in Appendix \ref{ap:conservation}, Equation (\ref{rhogx4b}) still guarantees conservation of total linear momentum, despite switching to $\rho_{i,j,k}$.  In all cases, round-off error momentum conservation requires discretized Poisson equation solutions accurate to machine precision (possible via FFTs, for example).

In this work, we have argued that our momentum source term is fourth order accurate; however, Equation (\ref{rhogx4b}) is actually only of third order accuracy in space since it is constructed from the numerical divergence of a fourth order accurate gravitational stress tensor.  Despite this technicality, solutions obtained using this gravitational acceleration are still of fourth order accuracy in space when the algorithm is coupled to a fourth order accurate (or higher) explicit, spatial \textit{and} temporal algorithm, wherein the source term multiplies the numerical timestep which itself is proportional to the cell width.  In analogy to pure hydrodynamics, only a fourth order accurate cell-surface-averaged pressure is required to achieve fourth order accurate solutions, despite the scheme having only a third order accurate pressure gradient force, when coupled to an otherwise fourth order accurate spatial and temporal integrator.  We demonstrate recovery of expected convergence rates in \S4 and omit any additional caveats on the accuracy of our algorithm.

\subsection{Gravitational Energy Release}

The high order gravitational energy release is 
\begin{widetext}
\begin{eqnarray}
\left[ \rho v _x g _x\right] _{i,j,k} & = &
\frac{1}{24} \left[ 
- \left( \rho v _x \right) _{i+\frac{3}{2},j,k} g _{x,i+\frac{3}{2},j,k} ^{(4)} + 13 \left( \rho v _x \right) _{i+\frac{1}{2},j,k} g _{x,i+\frac{1}{2},j,k} ^{(4)} + 13 \left( \rho v _x \right) _{i-\frac{1}{2},j,k} g _{x,i-\frac{1}{2},j,k} ^{(4)}  - \left( \rho v _x \right) _{i-\frac{3}{2},j,k} g _{x,i-\frac{3}{2},j,k} ^{(4)} 
\right] \nonumber \\
&& + \frac{1}{96} \left[ \left( \rho v _{x} \right) _{i+\frac{1}{2},j+1,k}- \left( \rho  v _x \right) _{i+\frac{1}{2},j-1,k} \right] \left[ g _{x,i+\frac{1}{2},j+1,k} ^{(4)} - g _{x,i+\frac{1}{2},j-1,k} ^{(4)} \right] \nonumber \\
&& + \frac{1}{96} \left[ \left( \rho v _{x} \right) _{i-\frac{1}{2},j+1,k} - \left( \rho v _x \right) _{i-\frac{1}{2},j-1,k}\right] \left[ g _{x,i-\frac{1}{2},j+1,k} ^{(4)} - g _{x,i-\frac{1}{2},j-1,k} ^{(4)} \right] \nonumber \\
&& + \frac{1}{96} \left[ \left( \rho v _{x} \right) _{i+\frac{1}{2},j,k+1}- \left( \rho  v _x \right) _{i+\frac{1}{2},j,k-1} \right] \left[ g _{x,i+\frac{1}{2},j,k+1} ^{(4)} - g _{x,i+\frac{1}{2},j,k-1} ^{(4)} \right] \nonumber \\
&& + \frac{1}{96} \left[ \left( \rho v _{x} \right) _{i-\frac{1}{2},j,k+1} - \left( \rho v _x \right) _{i-\frac{1}{2},j,k-1}\right] \left[ g _{x,i-\frac{1}{2},j,k+1} ^{(4)} - g _{x,i-\frac{1}{2},j,k-1} ^{(4)} \right], \label{rhovg4x}
\end{eqnarray}
\end{widetext}
where $ [\rho v _y g _y ] _{i,j,k} $
and $ [\rho v _z g _z ] _{i,j,k} $ are obtained by replacing $(x, y, z)$
and $ (i, j, k) $ cyclically.  Importantly, the fourth order accurate cell-surface-averaged mass flux evolving the continuity equation should be used for evaluations of $\rho v_x$  \citep[see, e.g.,][]{mikami08,springel10,mullen21}.  Equation (\ref{rhovg4x}) takes account of the fourth order accurate mass flux and gravity, and their second derivatives in the $ x $-direction in the first line of the right hand side of Equation (\ref{rhovg4x}). The remaining terms on the right hand side denote the product of the gradient of the mass flux and gravity in the $y$- and $ z $-directions. Such gradients (i.e., those tangential to the cell surface) are also taken into account in PPM \citep{colella84}.

The gravitational energy release should coincide with the change in the gravitational energy.
As shown in \cite{mullen21}, we can evaluate the gravitational energy from the gravity by Equation (\ref{Egrav}).  Using the fourth order accurate gravity on the cell surface, we can estimate the gravitational energy as
\begin{eqnarray}
\varepsilon _{\rm G} & = & - \frac{h^3}{8\pi G}
\sum _i \sum _j \sum _k
\left\{ \left[ g _{x,i+\frac{1}{2},j,k} ^{(4)} \right] ^2 \right. \nonumber \\
&& \left. + \left[ g _{y,i,j+\frac{1}{2},k} ^{(4)}\right] ^2
+ \left[ g _{z,i,j,k+\frac{1}{2}} ^{(4)}\right] ^2 \right. \nonumber \\
& & + \frac{1}{48} \left[ \left( g _{x,i+\frac{1}{2},j+1,k} - g _{x,i+\frac{1}{2},j-1,k} \right) ^2 \right. \nonumber \\
&& \hspace{18pt} + \left( g _{x,i+\frac{1}{2},j,k+1} -g _{x,i+\frac{1}{2},j,k-1} \right) ^2  \nonumber \\
&& \hspace{18pt} + \left( g _{y,i+1,j+\frac{1}{2},k} - g _{y,i-1,j+\frac{1}{2},k} \right) ^2 \nonumber \\
&& \hspace{18pt} + \left( g _{y,i,j+\frac{1}{2},k+1} -g _{y,i,j+\frac{1}{2},k-1} \right) ^2 \nonumber \\
&& \hspace{18pt} + \left( g _{z,i+1,j,k+\frac{1}{2}} - g _{z,i-1,j,k+\frac{1}{2}} \right) ^2 \nonumber \\
&& \hspace{18pt} + \left. \left. \left( g _{z,i,j+1,k+\frac{1}{2}} -g _{z,i,j-1,k+\frac{1}{2}} \right) ^2 \right] \right\} . \label{Egrav4}
\end{eqnarray}
Equation (\ref{Egrav4}) takes account of the gradient of the gravity on each cell surface as in Equation (\ref{rhovg4x}). Equation (\ref{Egrav4}) is of fourth order accuracy in space. 

Equation (\ref{energy2}) indicates that total energy is conserved when $\mathbf{\nabla} \cdot \mathbf{F} _g$ = 0 holds in the discretized form. We can rewrite this condition as 
\begin{eqnarray}
\mathbf{\nabla} \cdot \left(\rho \mathbf{v} - \frac{1}{4\pi G}\frac{\partial \mathbf{g}}{\partial t} \right) & = & 0, \label{Econ1} \\
\mathbf{g} \cdot \left(\rho \mathbf{v} - \frac{1}{4\pi G}\frac{\partial \mathbf{g}}{\partial t} \right) & = & 0 . \label
{Econ2} 
\end{eqnarray}
Equation (\ref{dPoisson2}) is consistent with Equation (\ref{Econ1}), while Equations (\ref{rhovg4x}) and (\ref{Egrav4}) are consistent with Equation (\ref{Econ2}).

\section{Test Problems}
We next present a suite of test problems evaluating the accuracy and robustness of our proposed algorithm.  Many of these tests were also presented in HM24 and HM25; we here supplement them with additional details and additional datasets; e.g., in this work, the algorithm is separately implemented in \textit{both} (1) an in-house, experimental coding platform and (2) the \texttt{Athena++} framework \citep{stone20}.  We test a variety of time integrators, including the third order accurate Runge-Kutta integrator \citep[RK3, see e.g.,][]{shu88} and a fourth order accurate, strong-stability preserving (SSP), low-storage RK4 integrator \citep{ketcheson10}.  Both MP5 \citep{sh1997} and an extremum-preserving variant of PPM \citep{colella84,colella08} are used for reconstruction.

\subsection{Jeans Linear Waves}
\label{sec::jeans}
We consider the propagation of small amplitude, self-gravity-modified linear waves with an initial condition specified by
\begin{eqnarray}
\rho (\mathbf{r}) & = & \rho _0 \left( 1 + \varepsilon \sin \mathbf{k}\cdot\mathbf{r} \right) , \\
\mathbf{v} (\mathbf{r}) & = & \frac{\varepsilon \omega \mathbf{k}}{|\mathbf{k}|^2} \sin \mathbf{k} \cdot \mathbf{r} , \label{LWv} \\
p (\mathbf{r}) & = & p _0 \left( 1 + \varepsilon \gamma \sin \mathbf{k}\cdot\mathbf{r} \right) , \\
\omega & = & \sqrt{\displaystyle \frac{\gamma p _0}{\rho _0} |\mathbf{k} | ^2 - 4 \pi G \rho_0} ,\\
\mathbf{k} & = & (2, 4, 4) , \\
\mathbf{r} & = & (x, y, z), 
\end{eqnarray}
with $ \rho _0 = 1 $, $ p _0 = 3/5 $, $ \gamma = 5/3 $, $ 4 \pi G = 1 $, and $ \varepsilon = 10 ^{-6} $ (HM25).  Our computational volume has edge lengths $\left[ L_x, L_y, L_z \right] = 2\pi \left[ k_x^{-1}, k_y^{-1}, k_z^{-1} \right]$ and is resolved by $ N _x \times N _y \times N _z$ numerical cells (with $ N _y = N _z = N _x /2 $).  We test both MP5 and PPM and RK3 and RK4 (with timestepping invoking a CFL number equal to 0.3).

\begin{figure}
    \plotone{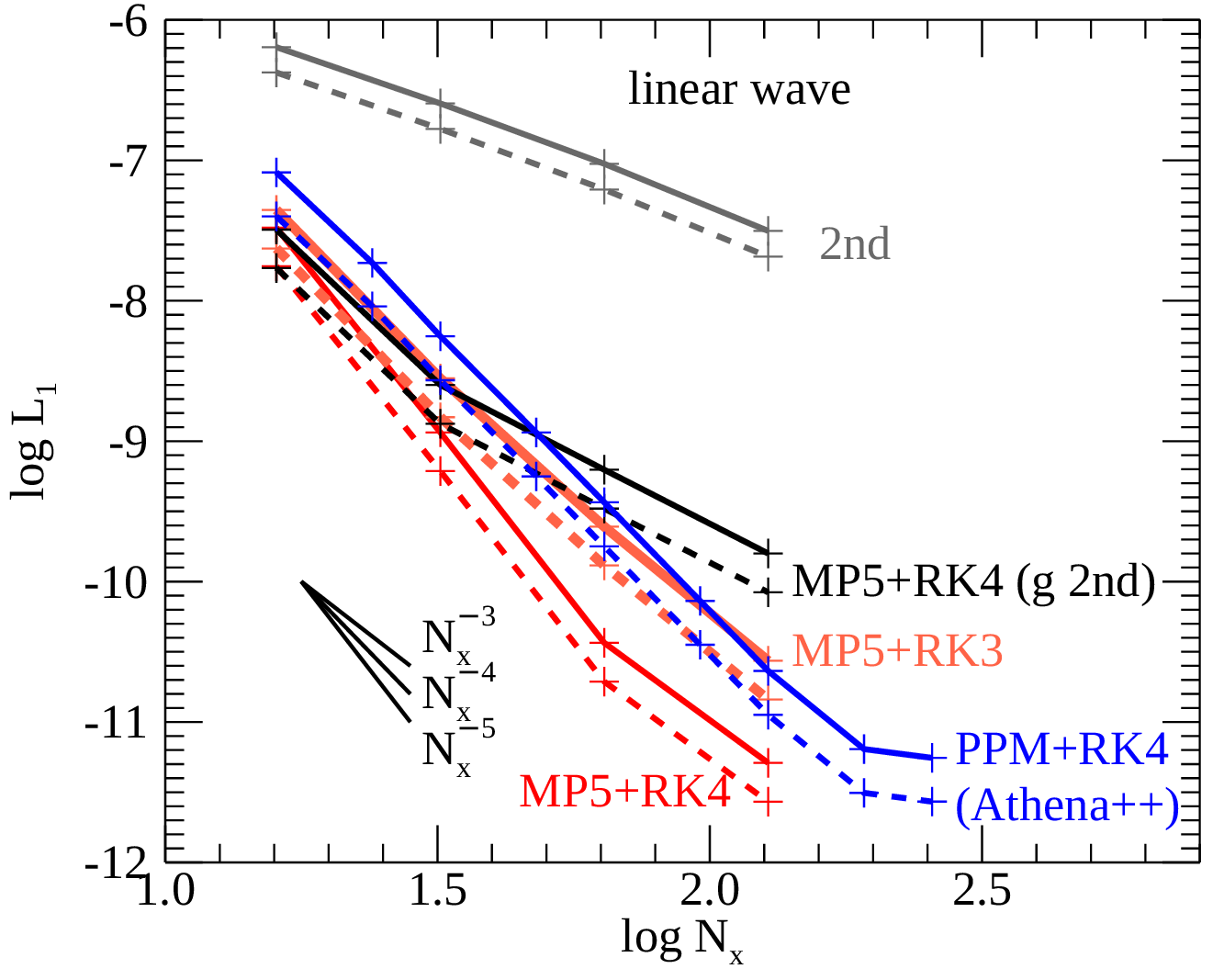}
    \caption{$L_1$ error convergence for self-gravity modified, small amplitude linear waves.}
    \label{fig:linear-w}
\end{figure}

Figure \ref{fig:linear-w} presents $L_1$ errors measured at $ t = 2 \pi/ \omega $ as a function of $N _x$.  Blue lines depict solutions obtained with \texttt{Athena++}.  Dashed lines denote the $L_1$ norm of the density, while solid ones report the sum of $L_1$ errors over all conserved variables. We test both second order gravity (gray and black lines) and fourth order accurate gravity (red and blue).  All algorithms invoking fourth order accurate gravity, fourth order accurate time integration, and high order reconstruction demonstrate at least fourth order convergence.  Even fifth order convergence ($\Delta \rho \propto N _{x} ^{-5} $) is demonstrated with MP5 reconstruction (thereby implying a dominance of spatial error in the hydrodynamics sector).  When second order gravity is coupled to high order hydrodynamics, only a second order convergence rate is observed beyond $N_x = 32$, thereby demonstrating the importance of our high order source terms.  A fully second order evaluation is presented in gray.

Note that $ L _1$ errors never drop below $ 3 \times 10 ^{-12} $ (i.e., $\sim$round-off).  Model convergence is additionally limited due to wave steepening, i.e., the nonlinearity of the wave propagation. The phase velocity of the Jeans linear wave is not constant in space, but a little higher where the density (and hence the sound speed) is higher. These effects are not taken account of in our reference analytic solution.  We have confirmed that the increase in $ L _1 $ errors due to nonlinearities is proportional to the square of the wave amplitude by repeating the tests with varying $ \alpha $. 

\subsection{Standing Wave Entropy Conservation}

Next, following HM25, we test the conservation of specific entropy via the evolution of standing waves. A uniform density and pressure medium (prescribed to $\rho_0$ and $p_0$ from \S \ref{sec::jeans}) is assigned a sinusoidal velocity (following Equation \ref{LWv}) with $ \varepsilon = 10 ^{-3} $.  With this configuration, the specific entropy is uniform across the domain in the initial condition.  The standing wave evolution should demonstrate amplitude steepening, all-the-while conserving specific entropy (until steepened into a shock).  We measure conservation of specific entropy by (1) converting fourth-order-accurate cell-volume-averaged conserved variables to fourth-order-accurate cell-centered-quantities via subtraction of the numerical Laplacian \citep[c.f.,][]{mccorquodale15,felker18}, (2) computing the cell-centered primitive vector from the cell-centered conserved vector, and (3) evaluating Equation (\ref{eq::spec_entr}) with the fourth-order-accurate cell-centered density and pressure normalized to $\rho_0$ and $p_0$, respectively.

\begin{figure}
\plotone{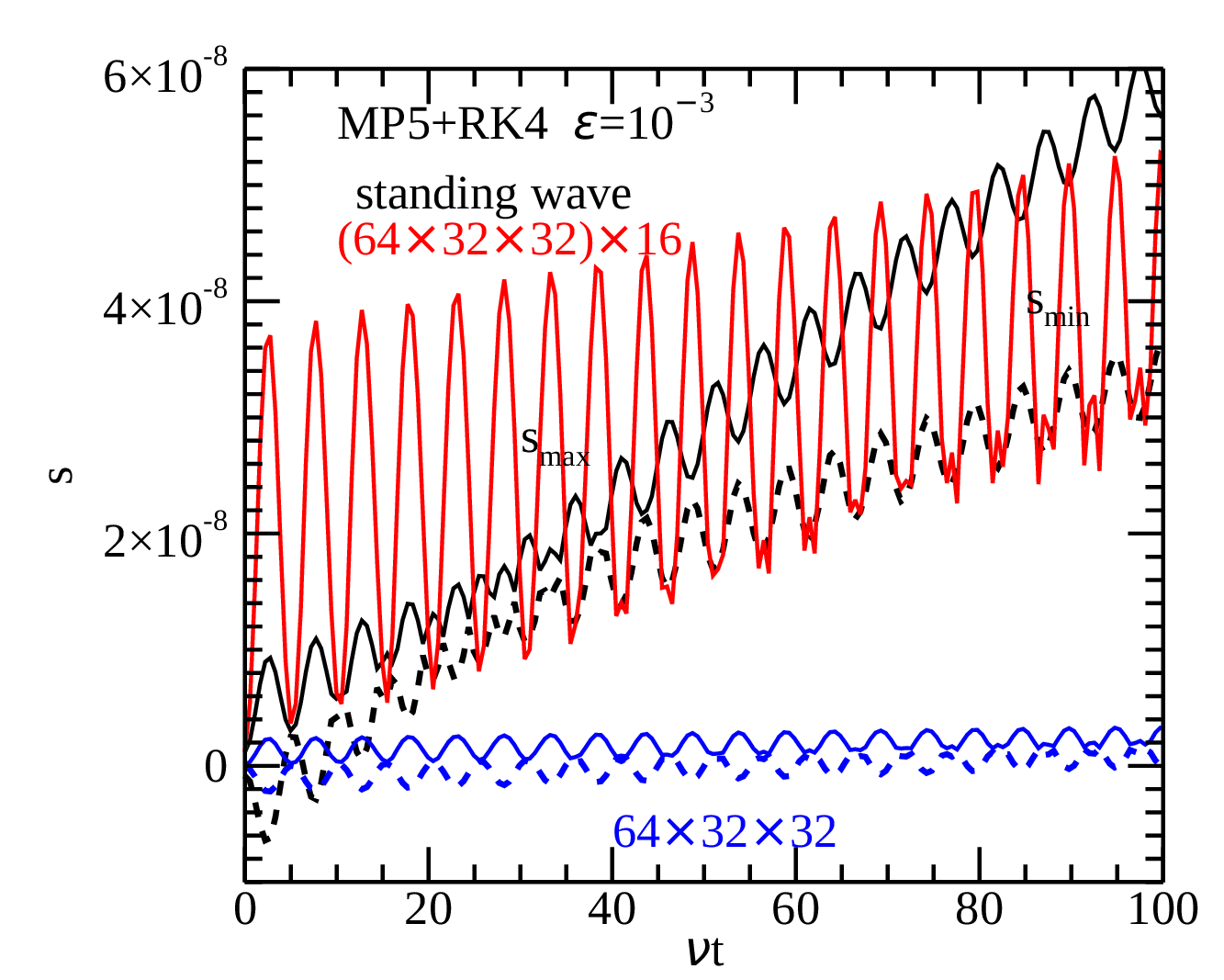}
\caption{Conservation of specific entropy in the standing wave problem. This figure was previously included in conference proceedings HM25.\label{fig:entropy-t}}
\end{figure}

Figure \ref{fig:entropy-t} presents the global maxima and minima of the specific entropy ($s _{\max}$, solid lines; $s _{\min} $, dashed lines; respectively) as a function of time (normalized by frequency $ \nu = \omega/2 \pi $) at two numerical resolutions, $[N _x, N _y, N _z] = [32,16,16] \; (\mathrm{black}) \; \mathrm{and} \; [64,32,32] \; (\mathrm{blue})$.  We overplot $ 16 {s} _{\max} $ for the $ [64,32,32] $ run to demonstrate 4th order convergence.

\begin{figure}
\plotone{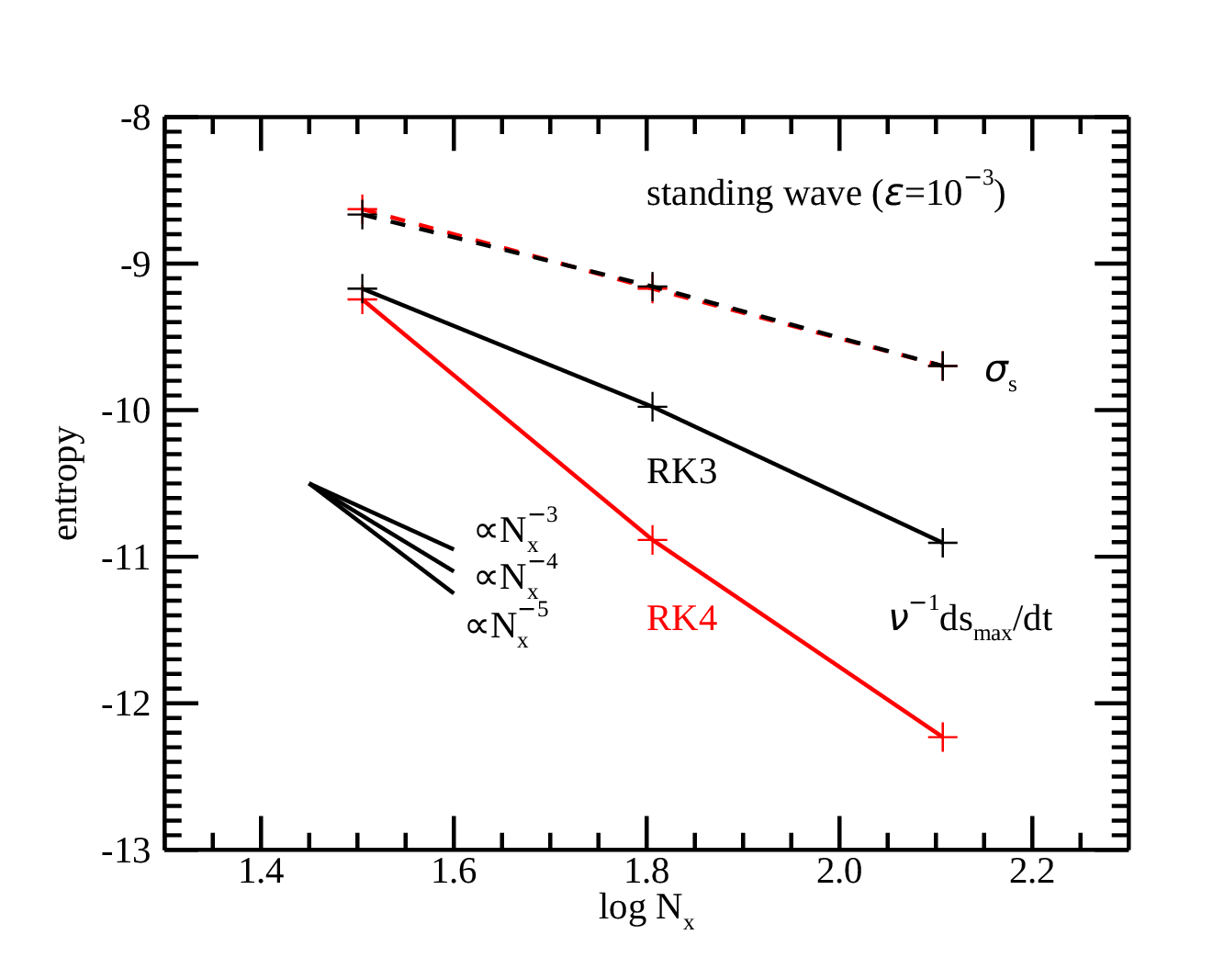}
\caption{Convergence of $\nu^{-1} d s_{\mathrm{max}} / dt$ and $\sigma_s$ for the standing wave test. This figure was previously included in conference proceedings HM25. \label{fig;entropy-N}}
\end{figure}

As in HM25, we perform linear fits to
\begin{eqnarray}
\tilde{s} _{\rm max} (t) & = & \tilde{s}  _{\rm max, 0} + \frac{d \tilde{s}  _{\rm max}}{dt} \frac{\omega t}{2 \pi} . \label{regression}
\end{eqnarray}
to measure deviations away from specific entropy conservation.  The slope $ \nu ^{-1} d {s} _{\rm max}/dt $ measures numerical energy dissipation from gravitational to thermal; we also measure oscillation standard deviation $\sigma_s$ whose error is (a) independent of the method and (b) inherent to truncation error associated with the application of the numerical Laplacian in our analysis procedure above.  Figure \ref{fig;entropy-N} presents the convergence of $ \nu ^{-1} d \tilde{s} _{\rm max}/dt $ and $ \sigma _s $ as a function of $ N _x $ when invoking MP5 reconstruction.  RK3 and RK4 solutions demonstrate third-order and fifth-order convergence, respectively.  Both RK3 and RK4 solutions for $ \sigma _s $ fall atop each other, as expected.  

\subsection{Advection of Self-gravitating Slabs in Equilibrium}
Implementation errors in, e.g., the $\rho \mathbf{v} \cdot \mathbf{g}$ source term, may be masked in the small amplitude linear waves tests in \S \ref{sec::jeans}.  Equally important, we now consider the advection of large-amplitude, self-gravitating, equilibrium slabs.  Following HM25, we assign density, velocity, and pressure distributions following
\begin{eqnarray}
\rho (\mathbf{r}) & = &
\rho _0 \left[ 1 + \varepsilon \cos \left( \mbox{\boldmath$k$} \cdot \mathbf{r} \right) + \frac{\varepsilon ^2}{3} \cos \left( 2 \mbox{\boldmath$k$} \cdot \mathbf{r} \right) \right] , \label{GSrho} \\
\mathbf{v} \left(\mathbf{r} \right) & = & (0.8, 0.6, 0.0), \\
p (\mathbf{r}) & = & 
p _0 + \frac{4 \pi G \varepsilon \rho _0 {}^2}{|\mbox{\boldmath$k$}|^2} \left[ \left( 1 - \frac{\varepsilon^2}{12} \right) \cos \left( \mbox{\boldmath$k$} \cdot \mathbf{r} \right) \right. \nonumber \\ 
&& \hspace{5pt}+ \frac{\varepsilon}{3} \cos \left( 2 \mbox{\boldmath$k$} \cdot \mathbf{r} \right) 
+ \frac{\varepsilon^2}{12} \cos \left( 3 \mbox{\boldmath$k$} \cdot \mathbf{r} \right) \nonumber \\
&& \hspace{5pt} \left. + \frac{\varepsilon^3}{144} \cos \left( 4 \mbox{\boldmath$k$} \cdot \mathbf{r} \right)\right], \\
\mathbf{k} & = & (1/3, 2/3, 2/3), \\
\mathbf{r} & = & (x, y, z),
\label{GSP} 
\end{eqnarray}
under periodic boundary conditions. We assign $ \rho _0 = 1 $, $ p _0 =6 $, $ \gamma = 5/3 $, $ 4 \pi G = 1 $, $ \varepsilon = 0.3 $, and $\mathrm{CFL} \; \mathrm{number} = 0.3$.  As in \S \ref{sec::jeans}, our computational volume has edge lengths $[L_x, L_y, L_z] = 2 \pi [k_x^{-1}, k_y^{-1}, k_z^{-1}]$ and is resolved by $N_x \times N_y \times N_z$ cells with linear resolution $h$ constant in all directions.  

\begin{figure}
\plotone{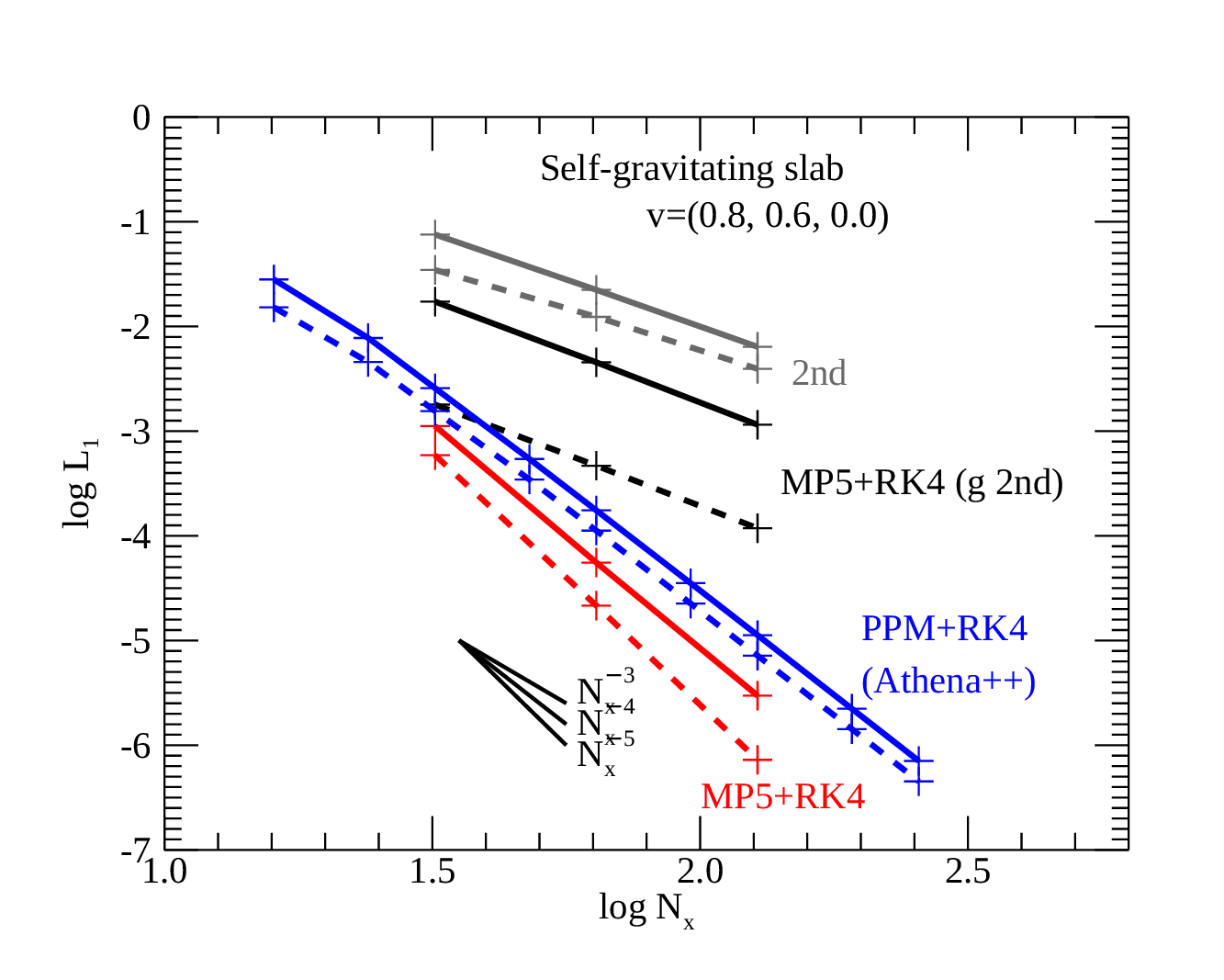}
\caption{Error in the moving slab problem.  The abscissa denotes the cell number in the $x$-direction while the ordinates denote the $L_1$ norm of the error in the conserved quantities (black) and the density (blue) at $ t = 2\pi /\mathbf{k}\cdot\mathbf{v} _0$.}
\label{fig:slab-L1-RK4}
\end{figure}

Figure~\ref{fig:slab-L1-RK4} presents $L_1$ errors measured at $t = 2\pi/(\mathbf{k} \cdot \mathbf{v})$ as a function of $ N _x $. Blue lines again depict solutions obtained with \texttt{Athena++}.  Again, all fully high order methods demonstrate at least fourth order convergence in the density $L_1$ error (dotted lines) and conserved vector $L_1$ error sum (solid lines), with lowest errors observed when using MP5 in conjunction with RK4 (red).  For these large amplitude slabs, the truncation error in the gravity is serious, as demonstrated by large errors associated with high order hydrodynamics coupled to second order gravity (black); a fully second order algorithm is shown for comparison (gray).

\begin{figure}
\plotone{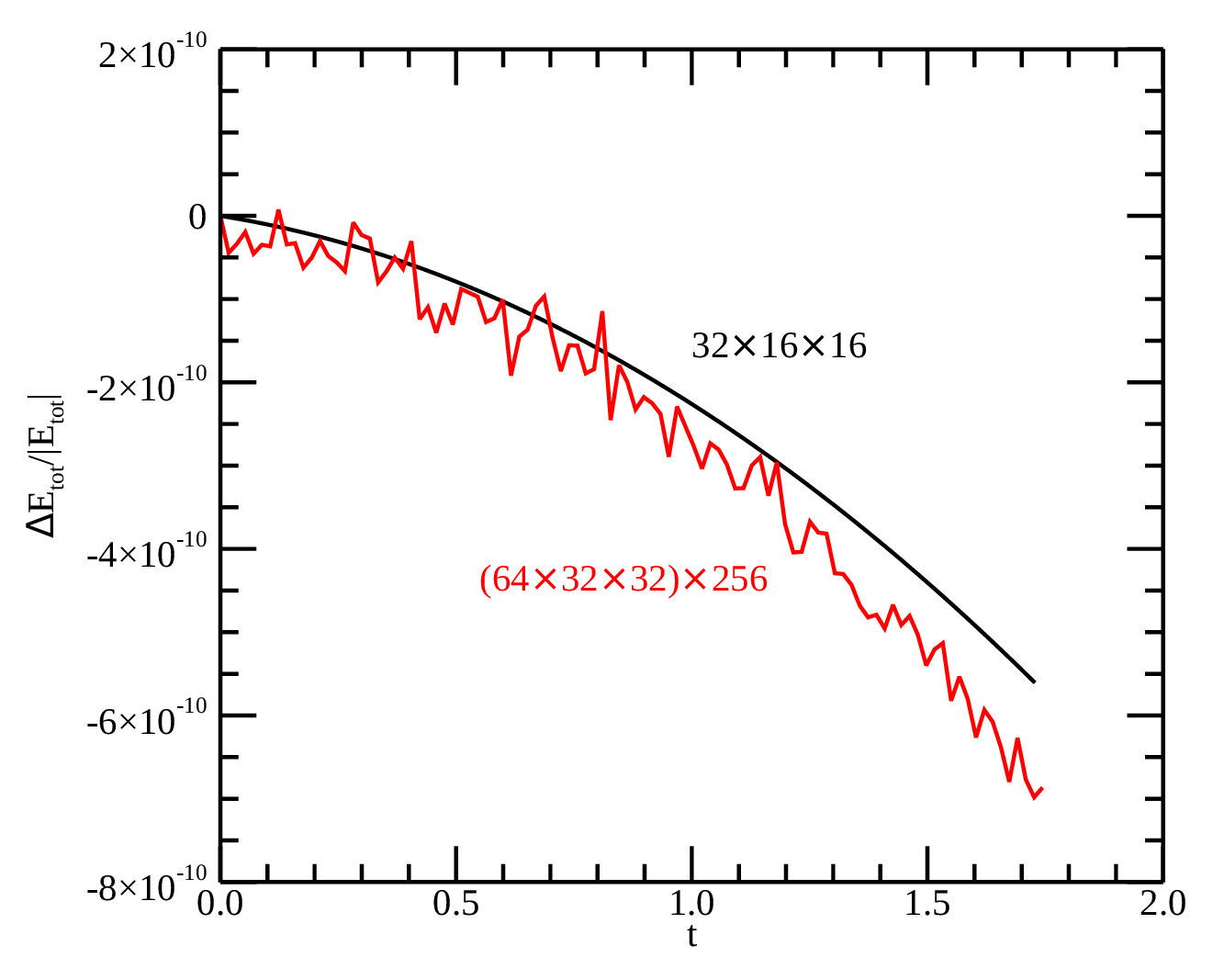}
\caption{The error in the total energy as a function of time $ t $ for the moving slab problem.  When $ \left( N _x, N _y, N _z \right) = (64, 32, 32)$, the error is a factor of 256 smaller than that for twice coarsened grid and only factor of 10 larger than numerical round-off error.}
\label{fig:slab-L1-energy}
\end{figure}

Global linear momentum conservation is confirmed to be at round-off error.  Figure~\ref{fig:slab-L1-energy} shows the temporal change in the total energy (including gravitational energy).  The black solid curve denotes the relative error in the total energy in the $ N _x = 32$ model.  The red solid curve denotes that for the $ N _x = 64 $ model after $256\times$ magnification.  Though associated with truncation, the error is comparable to round-off when $ N _x = 64 $.  

The $ N _x = 32 $ model has only 16 cells in the $y$- and $ z$-directions, while the density distribution contains a second overtone proportional to $ \cos 2 \mathbf{k} \cdot \mathbf{r} $.  Thus, the spatial resolution is quite low when $ N _x = 32$, yet still, the truncation error is small.

\subsection{$ n = 5$ Polytrope}
\label{sec::n5poly}

Next we consider $n=5$ polytropic equilibria, with a density, pressure, and gravitational potential specified by
\begin{eqnarray}
\rho (x, y, z) & = & \rho _{\rm c} \left( 1 + \frac{x^2 + y ^2 +z ^2}{3 R ^2} \right) ^{-5/2} ,  \\
p (x, y, z) & = & p _{\rm c}   \left( 1 + \frac{x^2 + y ^2 +z ^2}{3 R ^2} \right) ^{-3} ,  \\
\phi (x, y, z) & = & - 4 \pi G \rho _{\rm c} R \left( 1 + \frac{x^2 + y ^2 +z ^2}{3 R ^2} \right) ^{-\frac{1}{2}} ,
\end{eqnarray}
where $ \rho _{\rm c} $ and $p_c$ denote the central density and pressure, respectively.  As formulated, hydrostatic equilibrium is guaranteed when $ p _{\rm c} = 2 \pi G \rho _{\rm c} R $.  We assign model parameters $ \rho _{\rm c} = 1 $, scaling unit $ R = 1 $, $ 4 \pi G = 1 $, and adiabatic index $ \gamma = 5/3 $.

\begin{figure*}[ht]
\plotone{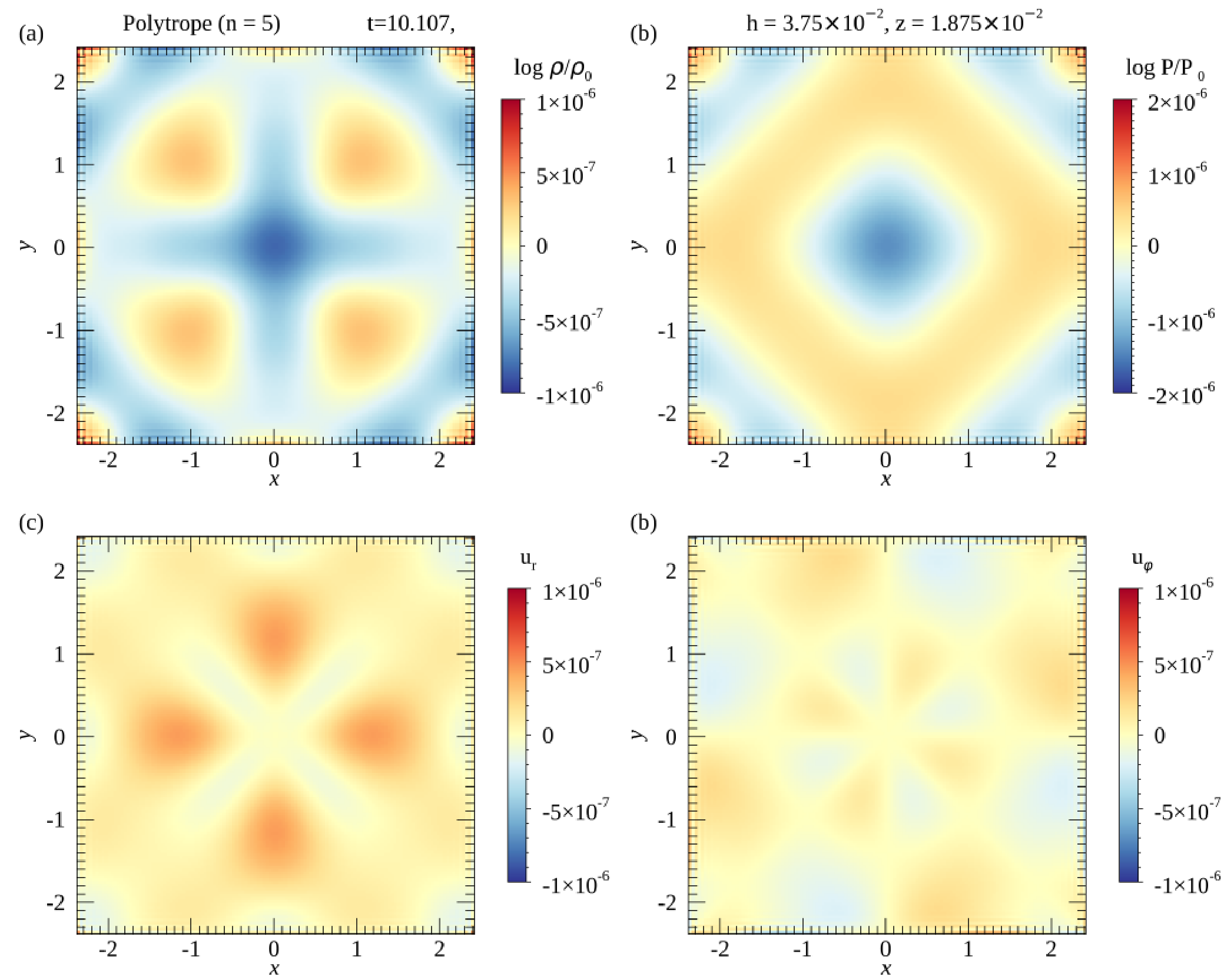}
\caption{Relative error in the density $\Delta \rho/\rho _0$, pressure $\Delta p/p _0$, radial velocity $u_r$ and azimuthal velocity $u_\varphi$ at $ t = 10.107 $ for the $n=5$ polytrope model. The associated movie shows its time evolution \color{red}{and its duration is 2 seconds. The relative error maintains quadrapolar symmetry, and wave reflections are notable at the outer boundary.} \label{fig:Pol5}}
\end{figure*}

Our computational volume is cubic with $4.8 R$ edge lengths resolved by 128$^3$ cells, corresponding to a linear resolution of $ h = 3.75 \times 10^{-3} R $.  We invoke MP5 reconstruction and an RK4 temporal integrator.  Figure~\ref{fig:Pol5} presents errors in the $ z = 1.875 \times 10 ^{-2} R$ plane after evolving to $ t = 10.107~\left(4\pi G \rho _{\rm c} \right)^{-\frac{1}{2}} $. Panels correspond to deviations from equilibrium values for the (upper-left) density, (upper-right) pressure, (lower-left) radial velocity and (lower-right) azimuthal velocity.  Careful numerical analysis is challenged by boundary conditions: we fix ``ghost cells" to their initial condition cell-volume-averaged conserved state vector values and mandate zero mass inflow; a dominant source of the error is associated with sound waves reflected off the outer boundary. 

\begin{figure}[ht]
\plotone{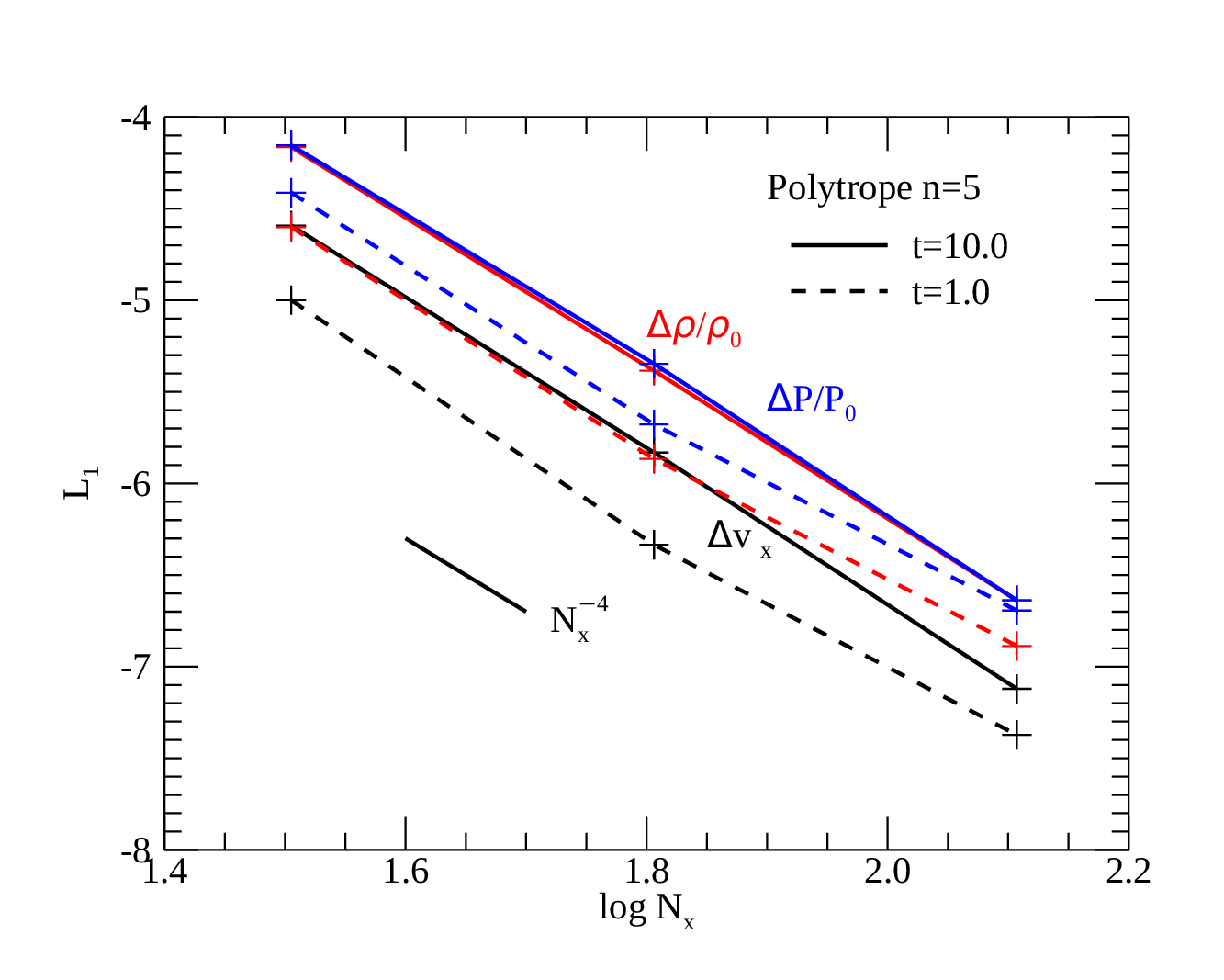}
\caption{Convergence test for the $n=5$ polytrope model. The abscissa denotes the cell number in the $x$-direction while the ordinates denotes $L_1$ norm of the error in $ \ln \rho $, $ \ln P $, and $ v _x $ at $ t = 1.0 $ (dashed) and 10.0 (solid). \label{fig:Pol5-con}}
\label{slab-L1-RK4}
\end{figure}

Figure~\ref{fig:Pol5-con} shows the convergence of errors. Lines denote $ \Delta \rho / \rho _0 $, $ \Delta p/p _0 $, and $ v _x $ $L_1$ errors at $ t = 1.0 \left(4\pi G \rho _{\rm c} \right)^{-\frac{1}{2}} $ (dashed) and $t = 10.0 \left(4\pi G \rho _{\rm c} \right)^{-\frac{1}{2}} $ (solid) .  Errors decrease in proportion to $ N _{x} ^{-4} $.

We achieve fourth-order accuracy in space even on the outer boundaries because we supplement the algorithm with analytic fourth-order accurate state vectors at the boundary.  We have examined this problem when not taking these special precautions, and as expected, the error is larger near the outer boundary.  We omit describing the details of these models, however, in most applications, we remark that outer boundary conditions may often be limited to second-order accuracy.

\subsection{$n =1$ Polytrope with Overlying Atmosphere \label{s+e}}

Next, following \cite{mullen21}, we consider the evolution of an $n=1$ polytrope with a light (but still self-gravitating), overlying atmosphere. The initial density, pressure, and mass profiles are expressed by
\begin{eqnarray}
\rho & = & 
\left\{
\begin{array}{lr}
\rho _{\rm c} \displaystyle \frac{\sin \alpha r}{\alpha r} & \text{if $r \le R $} \\
 \displaystyle \frac{M _{\rm a}}{2 \pi r ^5} & \text{otherwise,} \\
\end{array} \right. \\
p & = & \left\{ \begin{array}{ll}
\displaystyle \left( \rho _{\rm c} \frac{\sin \alpha r}{\alpha r} \right) ^2 & \text{if $r \le R$} \\
\displaystyle \left( \rho _{\rm c} \frac{\sin \alpha}{\alpha} \right) ^2 - \frac{G M _a R ^2}{48 \pi} \left[
\left( 4 M + M _{\rm a} \right) \right. & \\
\hskip 12pt \times \left. \displaystyle \left( \frac{1}{R^6} - \frac{1}{r ^6} \right) + \frac{3 Ma _{\rm a} R ^2}{r ^8}
\right] & \text{otherwise,} \\
\end{array} \right. \\
M & = & \int _0 ^R 4 \pi r ^2 \rho (r) dr \\ 
&  = & \frac{4 \pi \rho _{\rm c} \left( \sin \alpha R - \alpha R \cos \alpha R \right)}{\alpha ^3} , \\
M _{\rm a} & = & \varepsilon M, \\
r^2 & = & x ^2 + y ^2 + z ^2,
\end{eqnarray}
and we assign $ \rho _{\rm c} = 1 $, $ 4 \pi G =1 $, $ R = 1 $, $ \alpha = \sqrt{2 \pi} $, $ \varepsilon = 10 ^{-2} $, and adiabatic index $ \gamma = 5/3 $ as in \cite{mullen21}.

\begin{figure}[ht]
\plotone{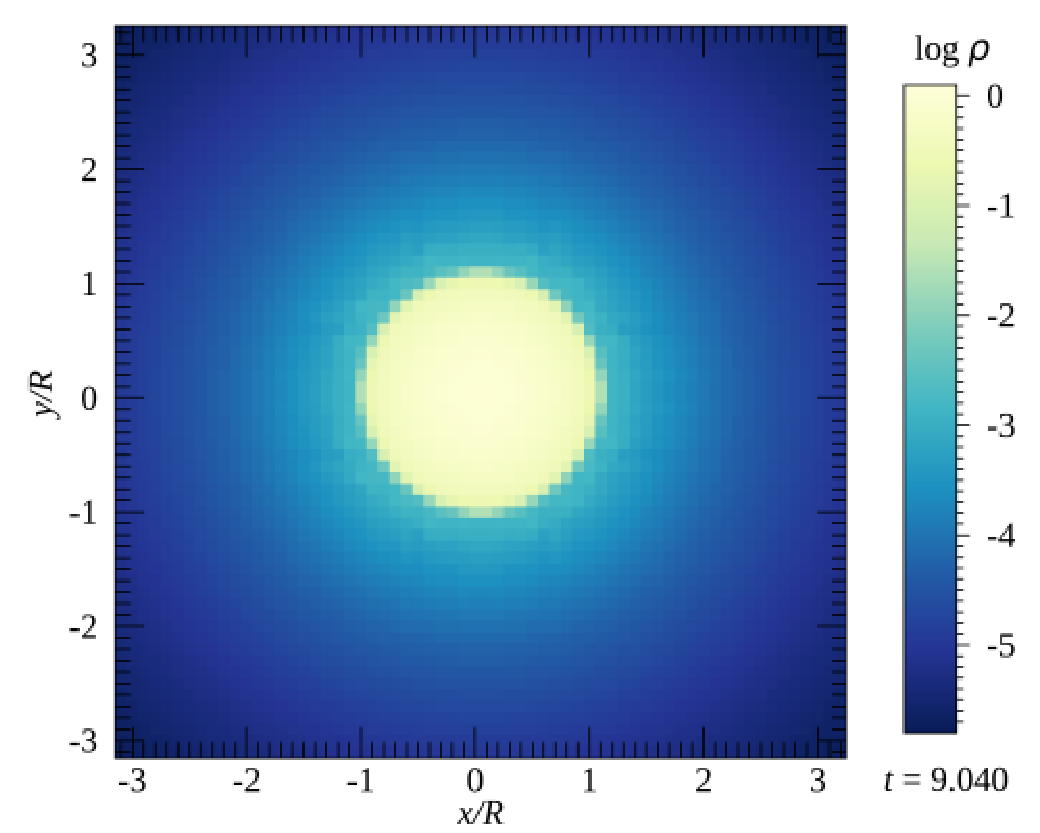}
\caption{$ n = 1 $ polytrope with a power law atmosphere. The color denotes the
the density distribution in the $ z = 0.05~R$ plane at $ t = 9.040~\left(4 \pi G \rho _{\rm c}\right)^{-\frac{1}{2}}$. 
The computational volume is resolved by $64^3$ cubic cells of linear resolution $h = 0.1~R$. The associated movie shows the density evolution from the initial condition to the final stage shown in the panel \color{red}{and its duration is 8 seconds. Since the change in the density is quite small, the apparent temporal evolution is minimal (as anticipated).} \label{fig:S+E}}
\end{figure}

Figure~\ref{fig:S+E} shows the density distribution at $ t = 9.040~\left(4\pi G \rho _{\rm c} \right)^{-\frac{1}{2}} $.  The equilibrium is sustained for more than 9 dynamical times.  No anomalous accelerations nor formation of plumes appear in our simulation \citep[as previously seen with $\nabla \times \mathbf{g} \neq 0$ discretizations of $\mathbf{T_g}$ in][]{mullen21}, though we do find appreciable change in the density and pressure near $ r = R $ (later discussed in \S5).

\subsection{Spherical Collapse}

Finally, following HM25, we evolve the ``inside-out" collapse  \citep[see, e.g.,][]{shu77} of an $n=5$ polytrope. The initial density and pressure profiles match those in \S \ref{sec::n5poly}, but with adiabatic index $\gamma = 1.3$.  We initiate gravitational collapse by manually setting the pressure interior to $ r < 9R/10$ to a uniform value.  Relative to \S \ref{sec::n5poly}, we extend the outer boundaries by $2\times$ and resolve the mesh with linear resolution $h \simeq ~0.08~R$.  

\begin{figure*}[ht]
\plottwo{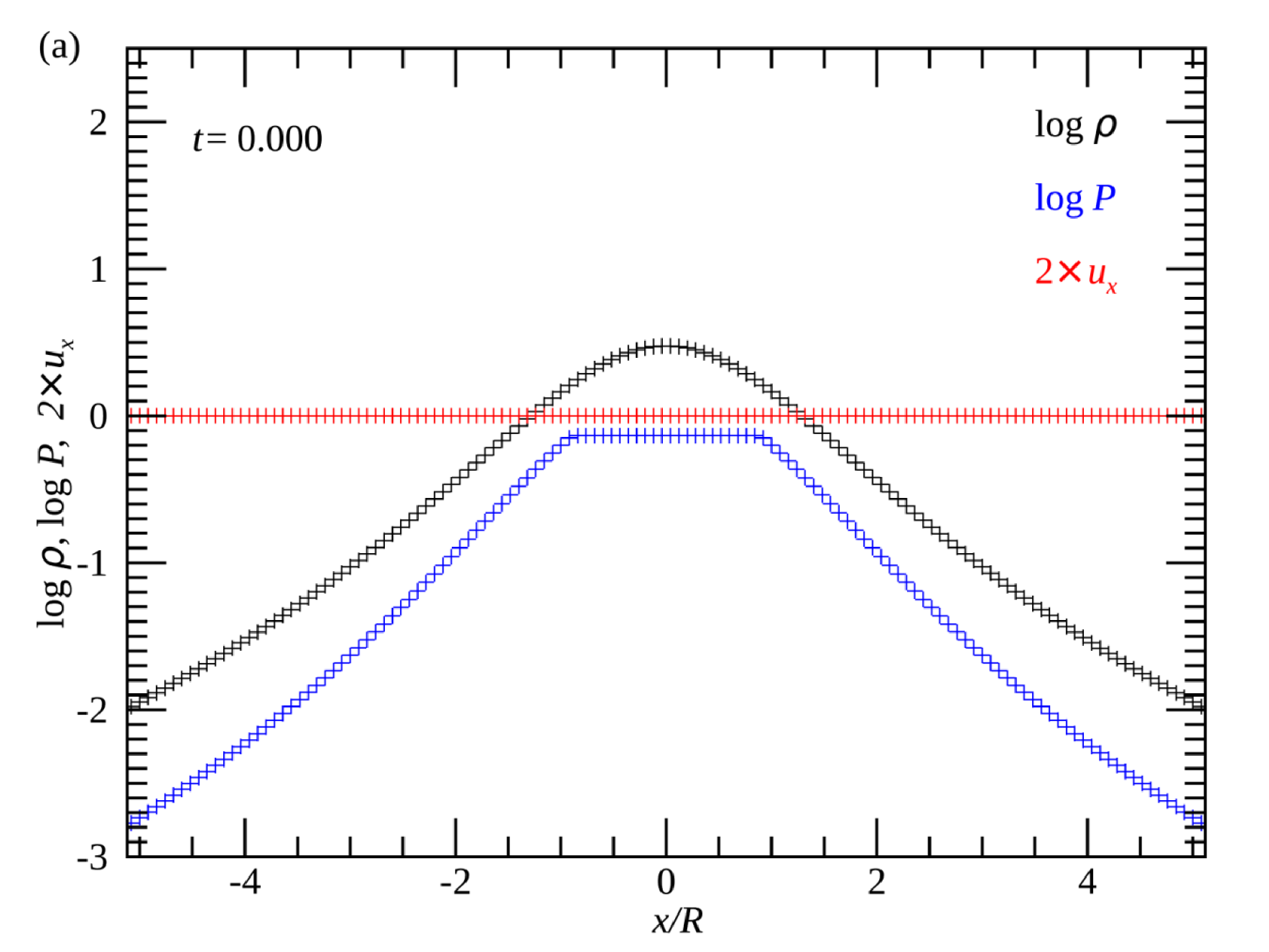}{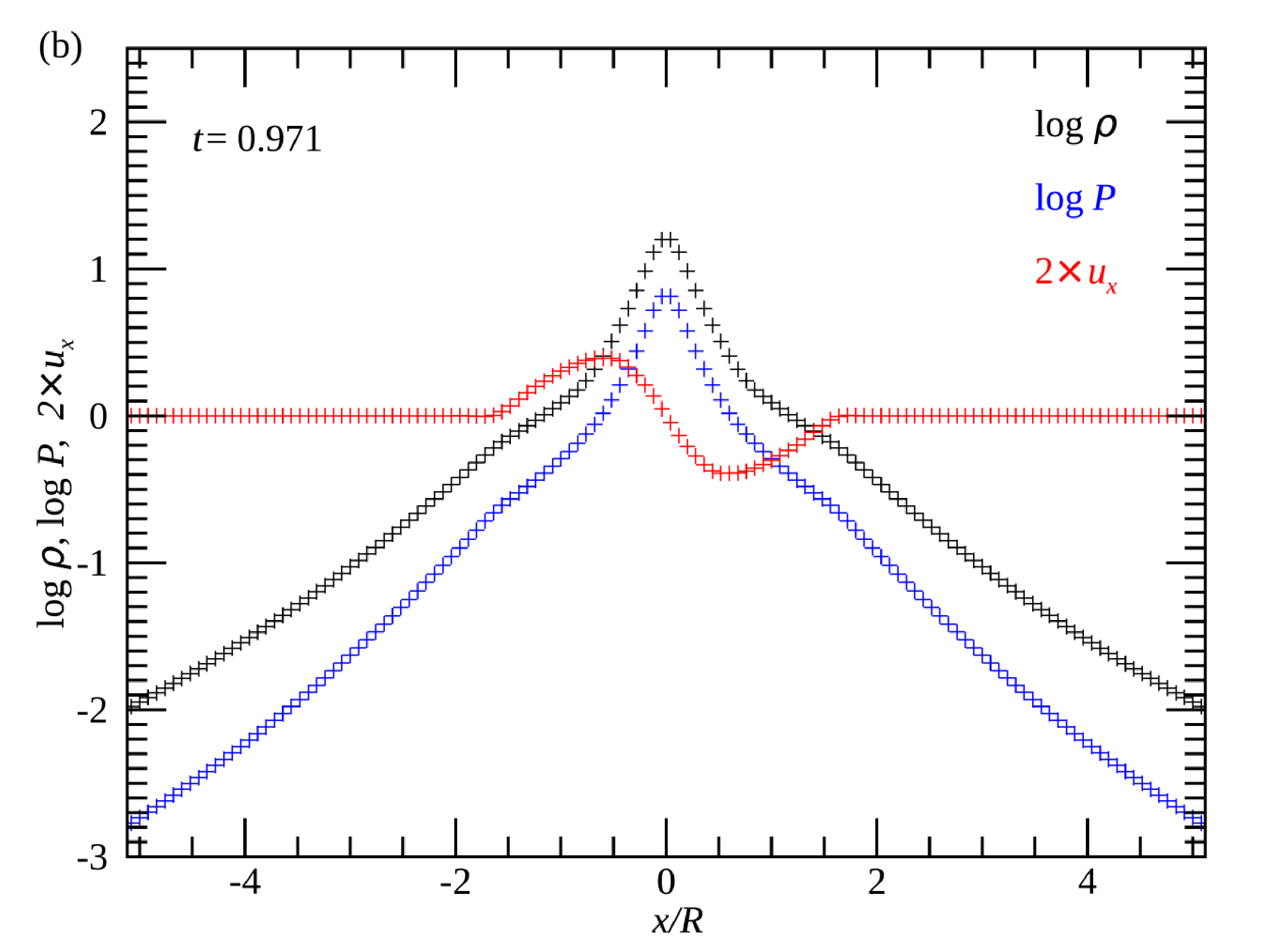}
\plottwo{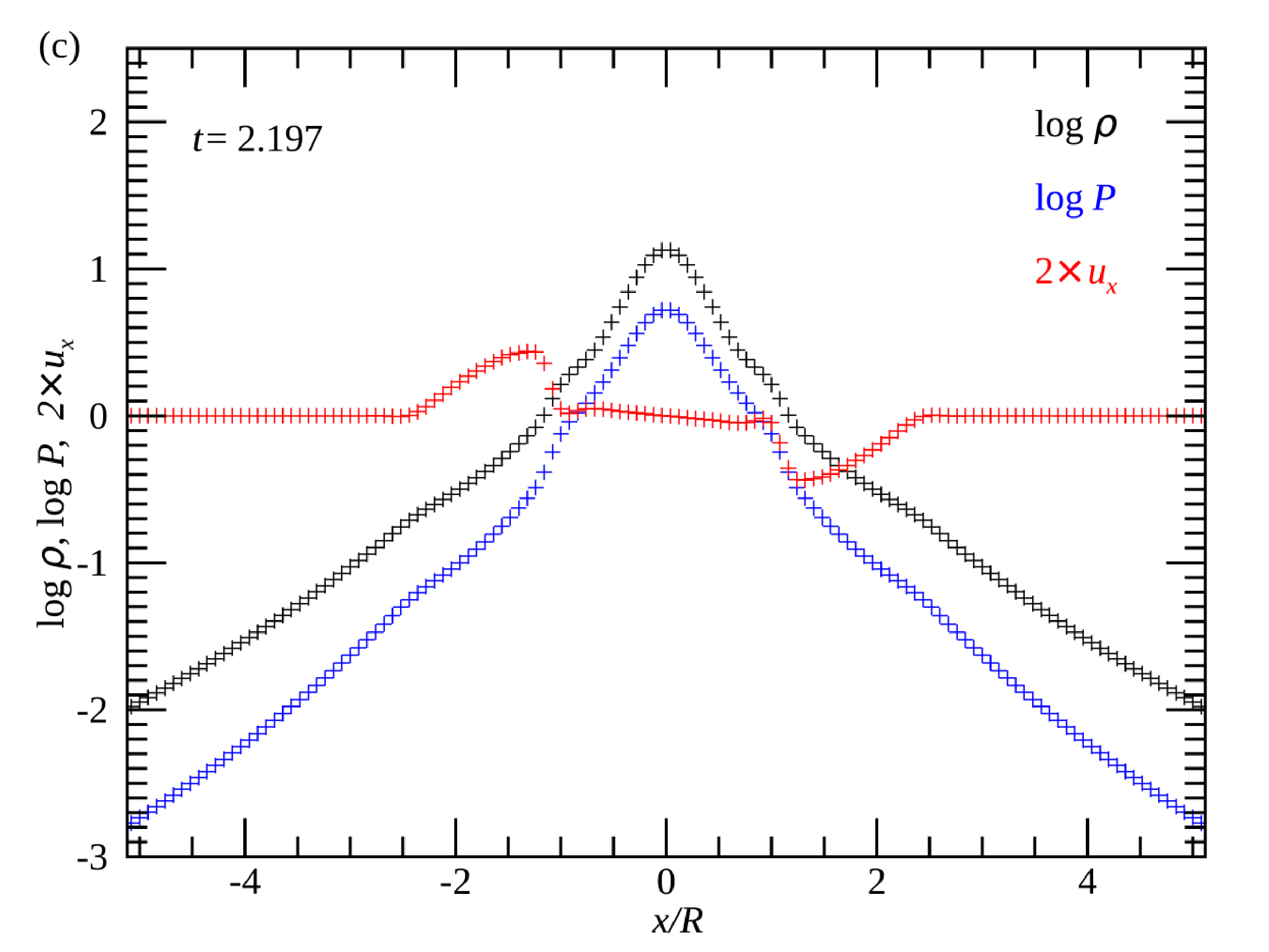}{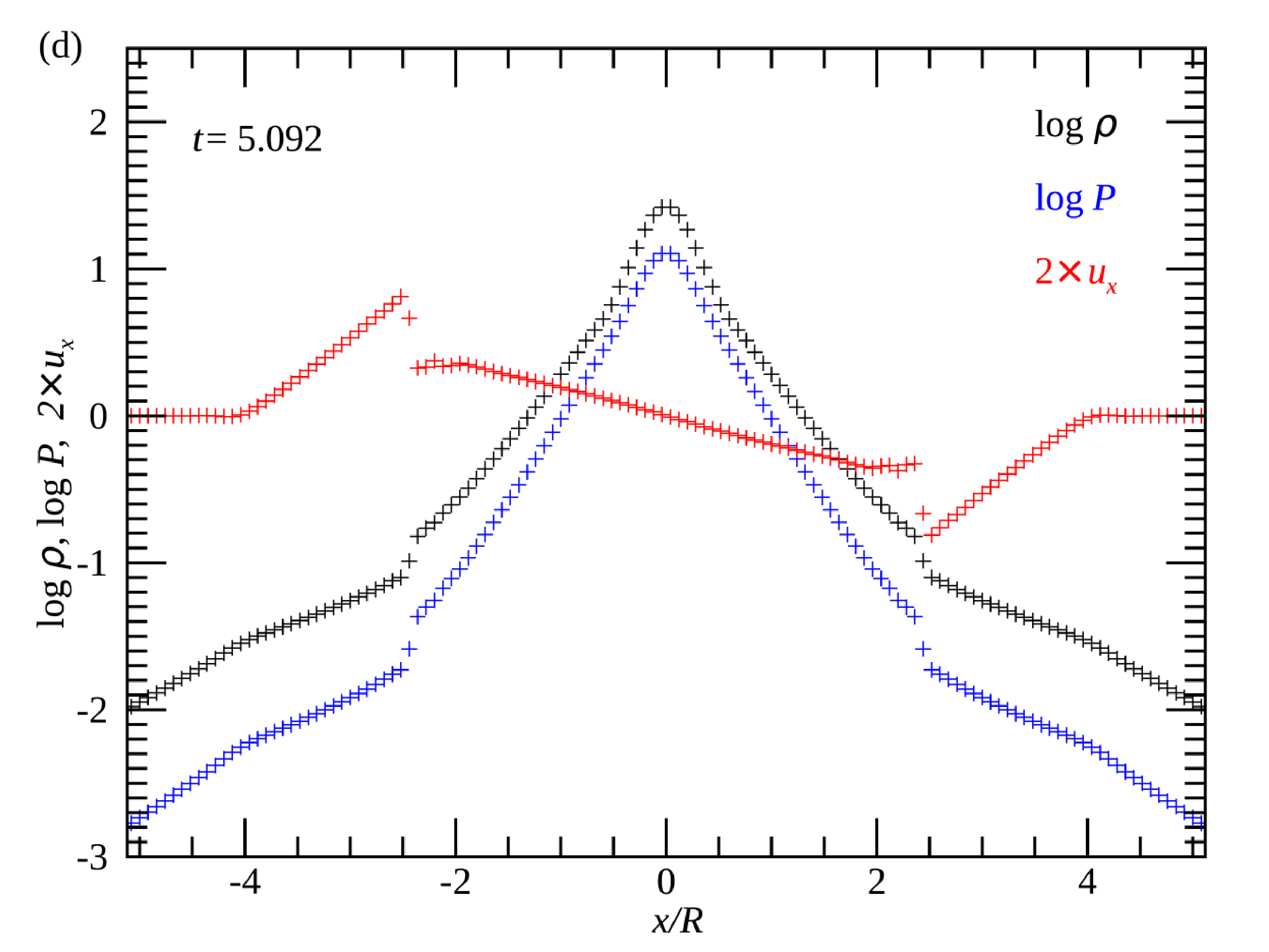}
\caption{Lineouts showing snapshots of the 3D collapse of an $n=5$ polytrope.  Black, blue and red crosses denote $\rho $, $ P $, and $ u _x $ along $ y = z = 0.04~R$. The associated movie shows the time evolution of the profiles \color{red}{and its duration is 3 seconds. The movie shows the formation and growth of a quasi-static core. It also shows the initiation of a second collapse of the core and the propagation of an accreting shock wave.}  This figure was previously included in conference proceedings HM25. 
\label{fig:3Dcollapse}}
\end{figure*}

The line-outs ($ y = z = 0.04~R$) in Figure~\ref{fig:3Dcollapse} denote the primitive state vectors (density $\rho$, black; pressure $p$, blue; and velocity $2 u_x$, red) as a function of $x$ at four times (0, upper-left; 0.971, upper-right; 2.197, lower-left; 5.092, lower right) all normalized to $ \left( 4 \pi G \rho _{\rm c} \right) ^{-\frac{1}{2}} $.  Intrinsic to the ``inside-out" collapse, we see an inward collapse for $r < 1.5 R $ expanding outwards while outer regions remain in hydrostatic equilibrium.  Beyond $ t = 2.197 $, we observe that the innermost regions of the polytrope are in a quasi-equilibrium surrounded by a spherical shock.  This quasi-equilbrium is short-lived, again collapsing at $t=5.092$

This model demonstrates the robustness of our scheme amidst non-linearity (e.g., shocks). We expect total energy conservation of fourth order accuracy until the wave fronts reach the outer boundary (wherein our fixed boundary condition eventually spoils conservation).  As presented in HM25, at $t=5.092 \left( 4 \pi G \rho _{\rm c} \right) ^{-\frac{1}{2}} $, we find $| \Delta E_\mathrm{tot} | / | E_\mathrm{tot}|~=~3.05 \times 10 ^{-8} $ and $| \Delta E_\mathrm{tot} | / |E_\mathrm{tot}| = 3.95 \times 10 ^{-7} $ in (a) the nominal model in Figure \ref{fig:3Dcollapse} and (b) a model with coarsened linear resolution $ h~=~0.16~R $, respectively, hence implying a total energy conservation error $\propto h ^{3.7} $.

\section{Discussion}

The numerical examples shown in the previous section demonstrate that our new source terms provide fourth-order accurate solutions when coupled to fourth order hydrodynamics. We emphasize, however, that the density may be be discontinuous (as in the polytropic gas sphere with overlying atmosphere model \S\ref{s+e}); accordingly, the second derivative of the gravitational potential can also be discontinuous. In these regimes, Equation (\ref{rhogx4b}) may not be an ideal discretization.

The first term in the right hand side of Equation (\ref{rhogx4b}) denotes the product of the cell-volume-averaged density and the cell-volume-averaged gravitational acceleration. The remaining terms denote the inner product of the density gradient and spatial derivative of the gravitational acceleration. The latter may be less robust when the density distribution is discontinuous, i.e., when the density contrast is extremely large between adjacent numerical cells. That is, when $ \rho _{i,j,k} \ll \rho _{i+1,j,k} $, terms proportional to $ \left( \rho _{i+1,j,k} - \rho _{i,j,k} \right) $ will ``over-correct" $ \left( \rho \mathbf{g} \right) _{i,j,k} $.

\begin{figure}
    \plotone{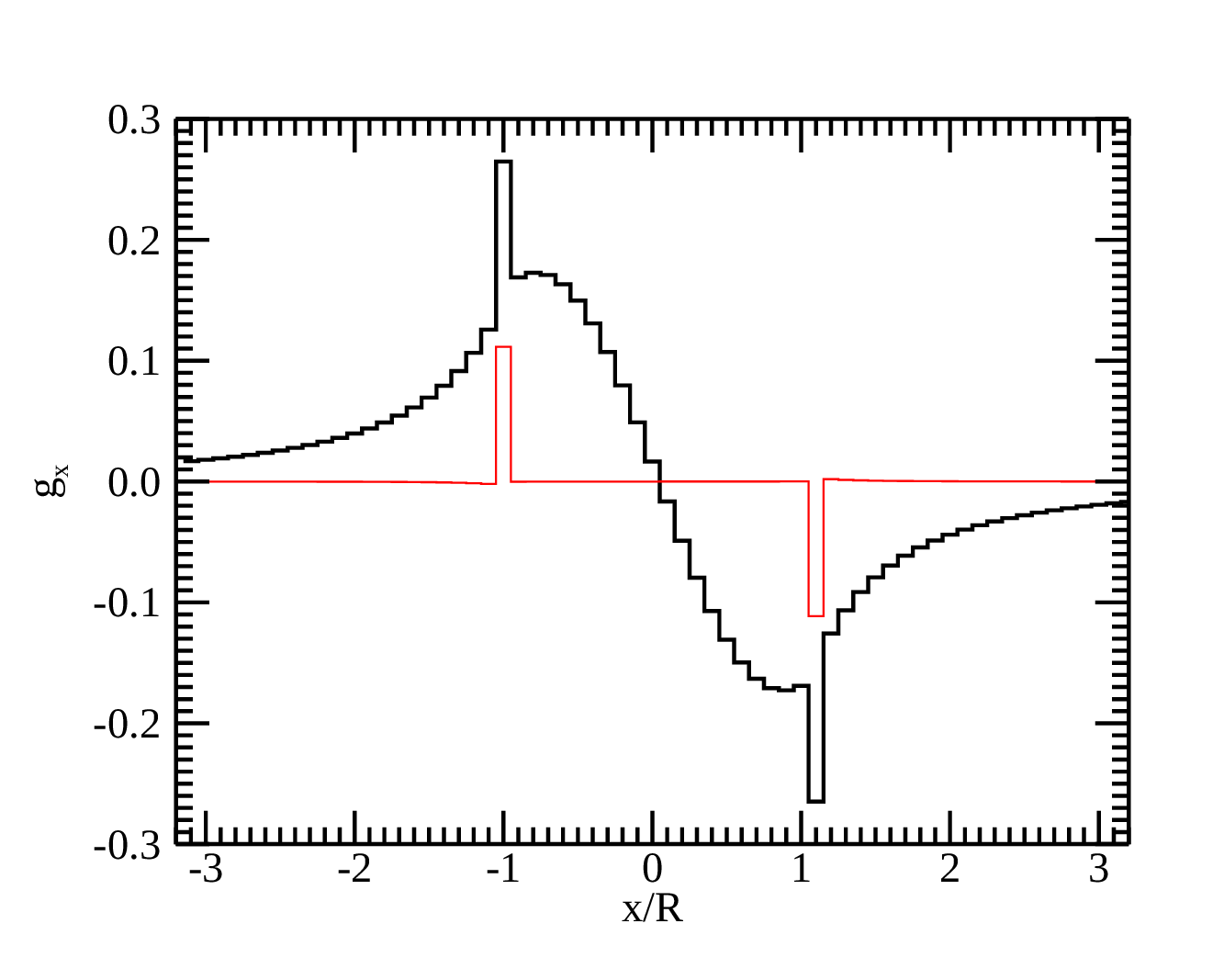}
    \caption{The black line denotes the gravity as a function of the cell center position
    in the cells along the line of $ y _j = z _j = 0.05~R$ for the polytrope of $ n = 1 $ with an envelope. The red line denotes the deviation from the cell average. \label{fig:gx}}
    \end{figure}

To illustrate the above argument, we examine in greater detail the gravitational acceleration for the $n=1$ polytrope with overlying atmosphere model in \S \ref{s+e}.  Figure \ref{fig:gx} shows a line-out of $ (\rho g _x) _{i,j,k} / \rho _{i,j,k} $ as a function of $ x _i$ for $ y _j = z _k = 5 \times 10 ^{-2}~R $ in black. The red line shows the deviation from the \textit{analytic} cell-volume-average $ \rho g _x $ to $ \rho $ ratio.  The deviation is large at $ x _i = \pm 1.05~R$ where the density contrast among adjacent cells is about 100$\times$.  

\begin{figure}
    \centering
    \plotone{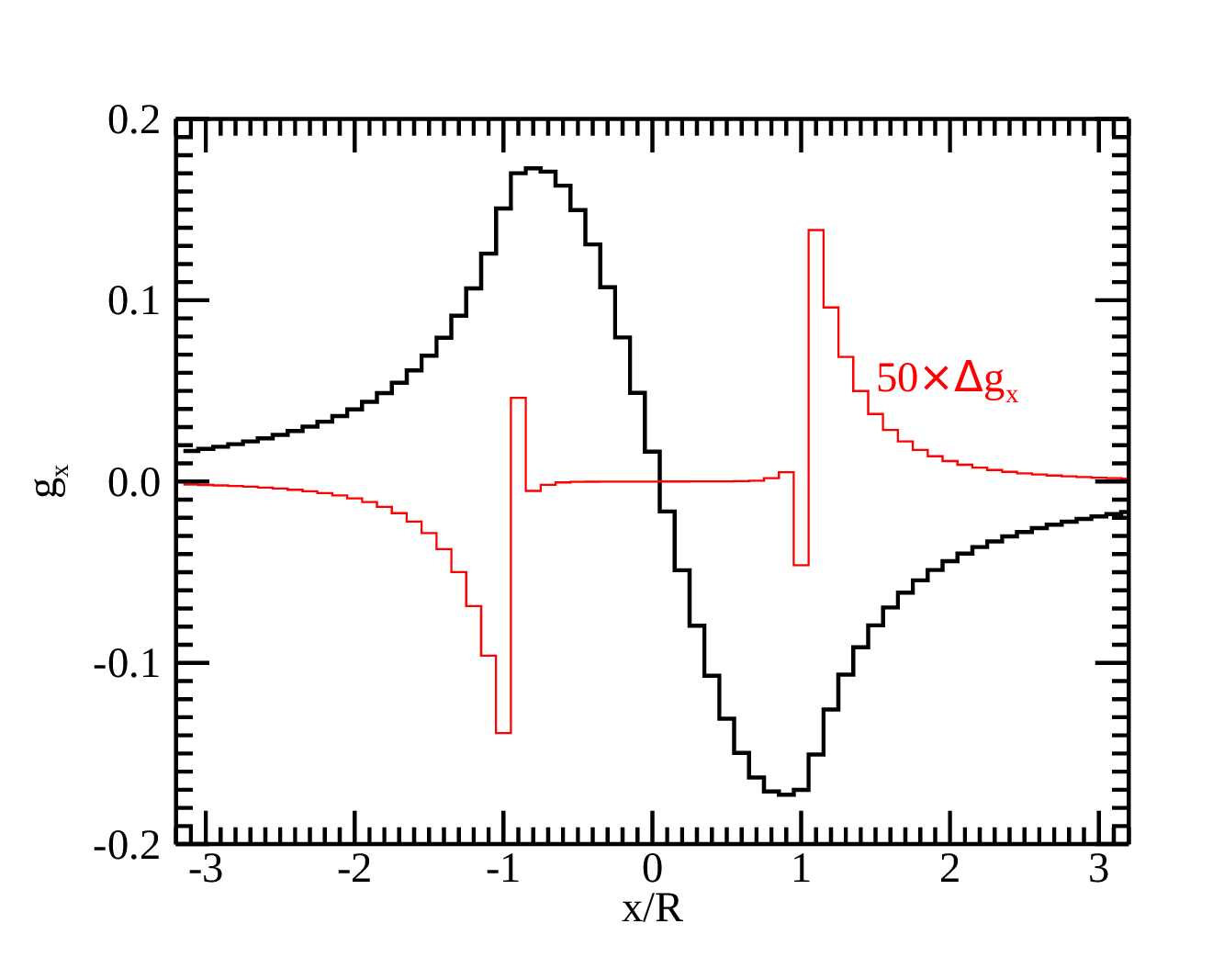}
    \caption{The same as Fig.~\ref{fig:gx} but after the large density gradient correction.
    The red line denotes the deviation magnified by a factor of 50.}
    \label{fig:gx2}
\end{figure}

When the density gradient is steep, suppression of the correction terms may improve the accuracy of the gravitational acceleration, however, modulating their magnitude alone may violate conservation of total linear momentum. Therefore, in order to conserve total linear momentum, we transfer the suppressed correction terms to the gravitational acceleration in the adjacent cell.  The correction terms should be suppressed only when the density contrast, e.g.,
\begin{eqnarray}
\Delta _{i+\frac{1}{2},j,k} & = &  \ln \rho _{i+1,j,k} / \rho _{i,j,k}  \label{rhocontrast} , 
\end{eqnarray}
exceeds a prescribed threshold.  In the following we set the threshold to be $ \left|\Delta _{i+\frac{1}{2},j,k} \right| > 0.5 $.  Consider the case when $ \Delta _{i+\frac{1}{2}, j,k} > 0.5 $, then we suppress the corresponding correction term in cell $(i,j,k)$   
by multiplying 
\begin{eqnarray}
f _{i+\frac{1}{2},j,k} ^{(i)} & = & 
1 - \tanh [5.0 (\Delta _{i+\frac{1}{2},j,k} - 0.5) ] , \label{gfactor}
\end{eqnarray}
and pass the suppressed correction term to the gravitational acceleration in cell $(i+1,j,k) $. Figure \ref{fig:gx2} is the same as Figure \ref{fig:gx} but after multiplying the factor given by Equation (\ref{gfactor}). The deviation from the true average is reduced greatly.  The magnitude and nature of the threshold suppression factor is arbitrary (and could potentially be improved), however, this is beyond the scope of this paper.

Similarly, we may need to modify Equation (\ref{rhovg4x}) since the mass flux can be discontinuous near shocks and contact discontinuities.  We can rewrite the first term in the right hand side of Equation (\ref{rhovg4x}) as
\begin{eqnarray}
& \hphantom{=} & 
\frac{1}{24} 
\left[
-\left( \rho v _x \right) _{i+\frac{3}{2},j,k} g _{x,i+\frac{3}{2},j,k} ^{(4)} \right. \nonumber \\
& & \left. + 13 \left( \rho v _x \right) _{i+\frac{1}{2},j,k} g _{x,i+\frac{1}{2},j,k} ^{(4)} \right. \nonumber \\
& & \left. + 13 \left( \rho v _x \right) _{i-\frac{1}{2},j,k} g _{x,i-\frac{1}{2},j,k} ^{(4)} 
 \left. - \left( \rho v _x \right) _{i-\frac{3}{2},j,k} g _{x,i-\frac{3}{2},j,k} ^{(4)} 
\right] \right. \nonumber = \\
& & \left.  \frac{1}{2} \left[ \left( \rho v _x \right) _{i+\frac{1}{2},j,k} g _{x,i+\frac{1}{2},j,k} ^{(4)}
 + \left( \rho v _x \right) _{i-\frac{1}{2},j,k} g _{x,i-\frac{1}{2},j,k} ^{(4)} \right] \right. \nonumber \\
& & \left.  - \frac{1}{24} \left[ \left( \rho v _x \right) _{i+\frac{3}{2},j,k} g _{x,i+\frac{3}{2},j,k} ^{(4)} - \left( \rho v _x \right) _{i-\frac{1}{2},j,k} g _{x,i-\frac{1}{2},j,k} ^{(4)} \right]  \right. \nonumber \\
& & \left.  + \frac{1}{24} \left[ \left( \rho v _x \right) _{i+\frac{1}{2},j,k} g _{x,i+\frac{1}{2},j,k} ^{(4)} - \left( \rho v _x \right) _{i-\frac{3}{2},j,k} g _{x,i-\frac{3}{2},j,k} ^{(4)} \right] \right. . \nonumber \label{rhovxgx-42}
\end{eqnarray}
The first line on the right hand side of the equality denotes the second order accurate gravitational energy release; the second and third lines denote 4th order correction terms evaluated on the cell surfaces. The correction terms include the mass flux and gravity across a cell surface outside the $(i,j,k)$ cell.  We omit these terms when the mass flux contains a large jump.  Each correction term appears in the gravitational release of the adjacent cells with opposite signs of each other.  Thus, total energy conservation remains fourth order accurate even when the correction terms are omitted.  The right hand side of Equation (\ref{rhovg4x}) has fourth order correction terms due to the gradient in the mass flux in the $ y $- and $ z $-directions.   These correction terms have the same sign between the adjacent cells.  Thus the truncation error in the total energy would become a second order small quantity if we suppressed them.  For this reason, we choose not to do so.

The mass flux may contain a large jump about contact discontinuities or large amplitude pressure contrasts.  Equation (\ref{gfactor}) can evaluate the strength of the contact discontinuity; similarly we can evaluate the amplitude of pressure contrasts via
\begin{eqnarray}
\Delta ^\prime _{i+\frac{1}{2},j,k} & = & \log \left( \frac{P _{i+1,j,k}}{P _{i,j,k}} \right)  .   
\end{eqnarray}
We can suppress energy flux correction terms by a factor of
\begin{eqnarray}
f _{i+\frac{1}{2},j,k} ^\prime & = & 1 - \tanh \left\{ 5.0 \left[ \max(|\Delta \prime_{i+\frac{1}{2},j,k} | - 0.5, 0.0)  \right] \right. \nonumber \\
& & \left. + \max(|\Delta ^\prime _{i+\frac{1}{2},j,k} | - 0.5, 0.0) \right\} .
\end{eqnarray}
Again, other threshold and suppression factors may perform better. 

Finally, we present arguments for when we expect our fourth order algorithm to significantly outperform the second order algorithm of \cite{mullen21}. Equation (\ref{dPoisson}), the discretized Poisson equation, is of fourth order accuracy in space.  When the typical size of a massive object is $ R $, truncation error is as small as $ (h/R) ^{-4} $. However, we expect the truncation error in the gravitational acceleration to be much larger than $ (h/R) ^{-4} $. Equation (\ref{rhogx4b}) motivates the need to care about the spatial variation of the density and gravity within a cell.  The correction term is as large as $ h ^2 R ^{-1} |\mathbf{\nabla} \ln \rho | $.  When the density scale height, $ |\mathbf{\nabla} \ln \rho| ^{-1}$, is shorter
than the typical scale of the system, this truncation error is larger than the truncation error in the gravitational potential.  Thus, the inclusion of the correction term in the gravitational acceleration will improve the simulation of outer stratified layers in a massive object.  This discussion is valid also in case that the gravity is external.  If the gravity is not uniform, we can reduce the truncation error in $ \rho \mathbf{g} $ and $ \rho \mathbf{v} \cdot \mathbf{g} $ by taking account of the spatial variation in $ \mathbf{g} $ according to Equations (\ref{rhogx4b}) and (\ref{rhovg4x}). The reduction in the truncation error should improve the solution greatly when the gas is cold, i.e., when the thermal energy is much smaller than the gravitational energy. 

\section{Conclusions}
In this work we have presented a fourth order accurate algorithm for self-gravitating hydrodynamics.  The algorithm supplies high order evaluations of the momentum source term $\rho \mathbf{g}$ and gravitational energy release $\rho \mathbf{v} \cdot \mathbf{g}$ given fourth order accurate solutions to the Poisson equation.  The momentum source term is derived from a fourth order accurate gravitational stress tensor whose particular discretization was motivated by high order Taylor expansions and the $\nabla \times \mathbf{g} = 0$ constraint on the gravity (see Appendices A and B for further details and caveats).  The energy source term is derived from the mass fluxes evolving the continuity equation and fourth-order accurate cell-surface-averaged gravity.  Together, these choices permit (i) global linear momentum conservation at round-off error (given round-off error accurate solutions to the discretized Poisson equation), (ii) global total energy conservation at fourth order accuracy, and (iii) local conservation of specific entropy at fourth order accuracy (in the absence of shocks).  

Our included test problems demonstrate that the algorithm (i)recovers expected fourth order convergence rates of $L_1$ errors, (ii) is devoid of the anomalous accelerations previously presented in \cite{mullen21}, and (iii) robustly integrates problems with mildly strong shocks.  Steep density and/or pressure contrasts may promote some challenge for the proposed algorithm; herein, we have presented several mitigation strategies, however, further development is necessary when considering problems much less smooth than those considered here.  

In applying combinations of RK3 and RK4 temporal integrators with PPM and MP5 reconstruction algorithms in both an in-house experimental code \textit{and} the \texttt{Athena++} framework \citep{stone20}, we have demonstrated that our algorithm is likely compatible with most pre-existing high-order hydrodynamics solvers; i.e., the only requirements are fourth-order-accurate mass fluxes and gravitational potentials.

Relative to its second-order-accurate counterpart in \cite{mullen21}, the additional computational expense accrued by our fourth order algorithm is mostly associated with the shift to high order hydrodynamics, not the gravity itself; i.e., our algorithm only mandates additional FLOPs in constructing the high order correction terms.  We do remark, however, that for common MPI-decompositions (as in \texttt{Athena++}), the mass fluxes must be either computed in an additional layer of ``ghost cells" or communicated. Additionally, we expect that our algorithm could be extensible to grids invoking block-structured, adaptive mesh refinement (AMR), assuming (i) the availability of fourth order accurate Poisson solutions on such grids; (ii) fourth order accurate prolongation and restriction operations; and (iii) special care surrounding flux corrections, since our gravitational energy release is dependent on the mass flux (see Equation(\ref{rhovg4x})).  Each of these subtopics warrants future work, but we do not otherwise anticipate that our algorithm imposes any additional ghost cell requirements relative to the hydrodynamics sector.

To arrive at our fourth order gravitational stress tensor, we applied a Taylor series expansion strategy that could equally be considered for high order extensions of (i) other dynamical equations (e.g., material strength---the deviatoric stress tensor), {(ii) hydrodynamics in curvilinear geometries, and/or even (iv) general relativistic (magneto)hydrodynamics. For example, a source term akin to Equation (\ref{rhogx4b}) may be identifiable for curvilinear geometries if one properly takes account of the non-uniform cell volume \citep[see, e.g.,][where high order cell surface reconstructions from cell-averaged conserved vectors are presented for non-uniform cylindrical and spherical grid]{mignone14}.  All in all, each of these prospects would require significant attention, and we leave them for future work.

\begin{acknowledgments}
This work was supported by JSPS KAKENHI Grant Numbers 19K03906, 20H00182, 20H05847. 

This work has been assigned a document release number LA-UR-24-33255.
\end{acknowledgments}

\vspace{5mm}

\software{Athena++ \citep{stone20}}

\appendix

\section{Fourth Order Corrections to the Gravitational Stress Tensor}
\label{ap:correction}

For brevity in the main text, we omitted detailed expressions for the fourth order gravitational stress tensor $\mathbf{T_g}^{(4)}$. However, finding $\mathbf{T_g}^{(4)}$ was instrumental to identifying the high order gravity presented in Equation (\ref{rhogx4b}) and ensuring total linear momentum conservation via equivalence (Equation \ref{divTg}).  For these reasons, we below detail the formulation of the fourth order gravitational stress tensor:
\begin{eqnarray}
T _{g,xx,i+\frac{1}{2},j,k} ^{(4)} & = & T _{g,xx,i+\frac{1}{2},j,k} +
\Delta T _{g,xx,i+\frac{1}{2},j,k} ^{(a)} + \Delta T _{g,xx,i+\frac{1}{2},j,k} ^{(b)} + 
\Delta T _{g,xx,i+\frac{1}{2},j,k} ^{(c)} + \Delta T _{g,xx,i+\frac{1}{2},j,k} ^{(d)} \nonumber \\
& & + \Delta T _{g,xx,i+\frac{1}{2},j,k} ^{(e)} \nonumber + \Delta T _{g,xx,i+\frac{1}{2},j,k} ^{(f)} , \label{Txx4s}\\
4 \pi G \Delta T _{g,xx,i+\frac{1}{2},j,k} ^{(a)} & = & - \left( \frac{\partial^2 \phi}{\partial x \partial y} \right)^2 \frac{h^2}{12} - \left( \frac{\partial ^2 \phi}{\partial x \partial z} \right)^2 \frac{h^2}{12} \\
& = &  - \frac{1}{24} \left( g _{x,i+\frac{1}{2},j+1,k} - g _{x,i+\frac{1}{2},j,k} \right) ^2
- \frac{1}{24} \left( g _{x,i+\frac{1}{2},j,k} - g _{x,i+\frac{1}{2},j-1,k} \right) ^2 \nonumber \\
& &  - \frac{1}{24} \left( g _{x,i+\frac{1}{2},j,k+1} - g _{x,i+\frac{1}{2},j,k} \right) ^2
- \frac{1}{24} \left( g _{x,i+\frac{1}{2},j,k} - g _{x,i+\frac{1}{2},j,k-1} \right) ^2 \label{tgxxa}\\
4 \pi G \Delta T _{g,xx,i+\frac{1}{2},j,k} ^{(b)} & = & - \frac{\partial \phi}{\partial x}
\frac{\partial ^3 \phi}{\partial x ^3} \frac{h^2}{12}\\
& = & - \frac{1}{12} g _{x,i+\frac{1}{2},j,k} \left( g _{x,i+\frac{3}{2},j,k} - 2 g _{x,i+\frac{1}{2},j,k} + g _{x,i-\frac{1}{2},j,k}  \right) , \\
4 \pi G \Delta T _{g,xx,i+\frac{1}{2},j,k} ^{(c)} & = & \left( \frac{\partial ^2 \phi}{\partial y^2} \right) ^2 \frac{h ^2}{12} +  \left( \frac{\partial ^2 \phi}{\partial z^2} \right) ^2 \frac{h ^2}{12}\\
& = & \frac{1}{12} \left( g _{y,i+1,j+\frac{1}{2},k} - g _{y,i+1,j-\frac{1}{2},k} \right)
\left( g _{y,i,j+\frac{1}{2},k} - g _{y,i,j-\frac{1}{2},k} \right) \nonumber \\
 & & + \frac{1}{12} \left( g _{z,i+1,j,k+\frac{1}{2}} - g _{z,i+1,j,k-\frac{1}{2}} \right)
\left( g _{z,i,j,k+\frac{1}{2}} - g _{y,i,j,k-\frac{1}{2}} \right) , \label{tgxxc} \\
4 \pi G \Delta T _{g,xx,i+\frac{1}{2},j,k} ^{(d)} & = &\frac{\partial \phi}{\partial y} \frac{\partial ^3 \phi}{\partial x^2 \partial y} \frac{h ^2}{6} + \frac{\partial \phi}{\partial z} \frac{\partial ^3 \phi}{\partial x^2 \partial z} \frac{h ^2}{6}\\
& = & \frac{1}{24} g _{y,i,j+\frac{1}{2},k} \left( g _{y,i+2,j+\frac{1}{2},k} - 2 g _{y,i+1,j+\frac{1}{2},k} + g _{y,i,j+\frac{1}{2},k} \right) \nonumber \\
& & + \frac{1}{24} g _{y,i,j-\frac{1}{2},k} \left( g _{y,i+2,j-\frac{1}{2},k} - 2 g _{y,i+1,j-\frac{1}{2},k} + g _{y,i,j-\frac{1}{2},k} \right) \nonumber \\
& & + \frac{1}{24} g _{y,i+1,j+\frac{1}{2},k} \left( g _{y,i+1,j+\frac{1}{2},k} - 2 g _{y,i,j+\frac{1}{2},k} + g _{y,i-1,j+\frac{1}{2},k} \right) \nonumber \\
&  & +\frac{1}{24} g _{y,i+1,j-\frac{1}{2},k} \left( g _{y,i+1,j-\frac{1}{2},k} - 2 g _{y,i,j-\frac{1}{2},k} + g _{y,i-1,j-\frac{1}{2},k} \right) \nonumber  \\
& & +\frac{1}{24} g _{z,i,j,k+\frac{1}{2}} \left( g _{z,i+2,j,k+\frac{1}{2}} - 2 g _{z,i+1,j,k+\frac{1}{2}} + g _{z,i,j,k+\frac{1}{2}} \right) \nonumber \\
& & + \frac{1}{24} g _{z,i,j,k-\frac{1}{2}} \left( g _{z,i+2,j,k-\frac{1}{2}} - 2 g _{z,i+1,j,k-\frac{1}{2}} + g _{z,i,j,k-\frac{1}{2}} \right) \nonumber \\
& & + \frac{1}{24} g _{z,i+1,j,k+\frac{1}{2}} \left( g _{z,i+1,j,k+\frac{1}{2}} - 2 g _{z,i,j,k+\frac{1}{2}} + g _{z,i-1,j,k+\frac{1}{2}} \right) \nonumber \\
&  & +\frac{1}{24} g _{z,i+1,j,k-\frac{1}{2}} \left( g _{z,i+1,j,k-\frac{1}{2}} - 2 g _{z,i,j,k-\frac{1}{2}} + g _{z,i-1,j,k-\frac{1}{2}} \right) , \\
4 \pi G \Delta T _{g,xx,i+\frac{1}{2},j,k} ^{(e)} & = & \frac{\partial \phi}{\partial y} \frac{\partial ^3 \phi}{\partial y^3} \frac{h^2}{6} +  \frac{\partial \phi}{\partial z} \frac{\partial ^3 \phi}{\partial z^3} \frac{h^2}{6}\\
& = & \frac{1}{24} g _{y,i+1,j+\frac{1}{2},k} \left( g _{y,i,j+\frac{3}{2},k} - 2 g _{y,i,j+\frac{1}{2},k} + g _{y,i,j-\frac{1}{2},k} \right) \nonumber \\
&& + \frac{1}{24} g _{y,i,j+\frac{1}{2},k} \left( g _{y,i+1,j+\frac{3}{2},k} - 2 g _{y,i+1,j+\frac{1}{2},k} + g _{y,i+1,j-\frac{1}{2},k} \right) \nonumber \\
&& + \frac{1}{24} g _{y,i+1,j-\frac{1}{2},k} \left( g _{y,i,j+\frac{1}{2},k} - 2 g _{y,i,j-\frac{1}{2},k} + g _{y,i,j-\frac{3}{2},k} \right) \nonumber \\
&&  +\frac{1}{24} g _{y,i,j-\frac{1}{2},k} \left( g _{y,i+1,j+\frac{1}{2},k} - 2 g _{y,i+1,j-\frac{1}{2},k} + g _{y,i+1,j-\frac{3}{2},k} \right) \nonumber \\
&& + \frac{1}{24} g _{z,i+1,j,k+\frac{1}{2}} \left( g _{z,i,j,k+\frac{3}{2}} - 2 g _{z,i,j,k+\frac{1}{2}} + g _{z,i,j,k-\frac{1}{2}} \right) \nonumber \\
&& + \frac{1}{24} g _{z,i,j,k+\frac{1}{2}} \left( g _{z,i+1,j,k+\frac{3}{2}} - 2 g _{z,i+1,j,k+\frac{1}{2}} + g _{z,i+1,j,k-\frac{1}{2}} \right) \nonumber \\
&& + \frac{1}{24} g _{z,i+1,j,k-\frac{1}{2}} \left( g _{z,i,j,k+\frac{1}{2}} - 2 g _{z,i,j,k-\frac{1}{2}} + g _{z,i,j,k-\frac{3}{2}} \right) \nonumber \\
&&  +\frac{1}{24} g _{z,i,j,k-\frac{1}{2}} \left( g _{z,i+1,j,k+\frac{1}{2}} - 2 g _{z,i+1,j,k-\frac{1}{2}} + g _{z,i+1,j,k-\frac{3}{2}} \right) , \label{tgxxe} \\
4 \pi G \Delta T _{g,xx,i+\frac{1}{2},j,k} ^{\rm (f)} & = & - \left( \frac{\partial ^2 \phi}{\partial y \partial z} \right) ^2 \frac{h^2}{12}\\ 
& = & - \frac{1}{48} \left( g _{y,i+1,j+\frac{1}{2},k+1} - g _{y,i+1,j+\frac{1}{2},k} \right)
\left( g _{y,i,j+\frac{1}{2},k+1} - g _{y,i,j+\frac{1}{2},k} \right) \nonumber \\
& & - \frac{1}{48} \left( g _{y,i+1,j-\frac{1}{2},k+1} - g _{y,i+1,j-\frac{1}{2},k} \right)
\left( g _{y,i,j-\frac{1}{2},k+1} - g _{y,i,j-\frac{1}{2},k} \right) \nonumber \\
& & - \frac{1}{48} \left( g _{y,i+1,j+\frac{1}{2},k} - g _{y,i+1,j+\frac{1}{2},k-1} \right)
\left( g _{y,i,j+\frac{1}{2},k} - g _{y,i,j+\frac{1}{2},k-1} \right) \nonumber \\
& & - \frac{1}{48} \left( g _{y,i+1,j-\frac{1}{2},k} - g _{y,i+1,j-\frac{1}{2},k-1} \right)
\left( g _{y,i,j-\frac{1}{2},k} - g _{y,i,j-\frac{1}{2},k-1} \right) \\
& = & - \frac{1}{96} \left( g _{y,i+1,j+\frac{1}{2},k+1} - g _{y,i+1,j+\frac{1}{2},k} \right)
\left( g _{y,i,j+\frac{1}{2},k+1} - g _{y,i,j+\frac{1}{2},k} \right) \nonumber \\
& & - \frac{1}{96} \left( g _{y,i+1,j-\frac{1}{2},k+1} - g _{y,i+1,j-\frac{1}{2},k} \right)
\left( g _{y,i,j-\frac{1}{2},k+1} - g _{y,i,j-\frac{1}{2},k} \right) \nonumber \\
& & - \frac{1}{96} \left( g _{y,i+1,j+\frac{1}{2},k} - g _{y,i+1,j+\frac{1}{2},k-1} \right)
\left( g _{y,i,j+\frac{1}{2},k} - g _{y,i,j+\frac{1}{2},k-1} \right) \nonumber \\
& & - \frac{1}{96} \left( g _{y,i+1,j-\frac{1}{2},k} - g _{y,i+1,j-\frac{1}{2},k-1} \right)
\left( g _{y,i,j-\frac{1}{2},k} - g _{y,i,j-\frac{1}{2},k-1} \right) \nonumber \\
& & - \frac{1}{96} \left( g _{z,i+1,j+1,k+\frac{1}{2}} - g _{z,i+1,j,k+\frac{1}{2}} \right)
\left( g _{z,i,j+1,k+\frac{1}{2}} - g _{z,i,j,k+\frac{1}{2}} \right) \nonumber \\
& & - \frac{1}{96} \left( g _{z,i+1,j+1,k-\frac{1}{2}} - g _{z,i+1,j,k-\frac{1}{2}} \right)
\left( g _{z,i,j+1,k-\frac{1}{2}} - g _{z,i,j,k-\frac{1}{2}} \right) \nonumber \\
& & - \frac{1}{96} \left( g _{z,i+1,j,k+\frac{1}{2}} - g _{z,i+1,j-1,k+\frac{1}{2}} \right)
\left( g _{z,i,j,k+\frac{1}{2}} - g _{y,i,j-1,k+\frac{1}{2}} \right) \nonumber \\
& & - \frac{1}{96} \left( g _{z,i+1,j,k-\frac{1}{2}} - g _{z,i+1,j-1,k-\frac{1}{2}} \right)
\left( g _{z,i,j,k-\frac{1}{2}} - g _{y,i,j-1,k-\frac{1}{2}} \right) , \label{Txx4e}
\end{eqnarray}
\begin{eqnarray}
T _{g,xy,i,j+\frac{1}{2},k} ^{(4)} & = & T _{g,xy,i,j+\frac{1}{2},k} ^{(2)} +
\Delta T _{g,xy,i,j+\frac{1}{2},k} ^{(a)} +
\Delta T _{g,xy,i,j+\frac{1}{2},k} ^{(b)} +
\Delta T _{g,xy,i,j+\frac{1}{2},k} ^{(c)} +
\Delta T _{g,xy,i,j+\frac{1}{2},k} ^{(d)} \nonumber \\ & & 
+ \Delta T _{g,xy,i,j+\frac{1}{2},k} ^{(e)}, \label{Txy4s} \\
4 \pi G \Delta T _{g,xy,i,j+\frac{1}{2},k} ^{(a)} & = &
\frac{\partial ^2 \phi}{\partial x^2} \frac{\partial ^2 \phi}{\partial x \partial y} \frac{h ^2}{12} \\
& = & \frac{1}{48} \left( g _{x,i+\frac{1}{2},j+1,k} + g _{x,i+\frac{1}{2},j,k} - g _{x,i-\frac{1}{2},j+1,k} - g _{x,i-\frac{1}{2},j,k} \right) \nonumber \\
& & \times \left( g _{y,i+1,j+\frac{1}{2},k} - g _{y,i-1,j+\frac{1}{2},k} \right) , \label{tgxya} \\
4 \pi G \Delta T _{g,xy,j+\frac{1}{2},k} ^{(b)} & = & - \frac{\partial \phi}{\partial x} \frac{\partial ^3 \phi}{\partial y ^3} \frac{h ^2}{12} \\
& = & \frac{1}{48}
\left( g_{x,i+\frac{1}{2},j+1,k} + g _{x,i-\frac{1}{2},j+1,k} + g _{x,i+\frac{1}{2}, j, k} + g _{x,i-\frac{1}{2},j,k} \right) \nonumber \\
& & \times \left( g _{y,i,j+\frac{3}{2},k} - 2 g _{y,i,j+\frac{1}{2},k} + g _{y,i,j-\frac{1}{2},k} \right), \\
4 \pi G \Delta T _{g,xy,i,j+\frac{1}{2},k} ^{(c)} & = & - \frac{\partial \phi}{\partial y} \frac{\partial ^3 \phi}{\partial x \partial y ^2} \frac{h ^2}{6} \nonumber \\
& = & - \frac{1}{24} g _{y,i,j+\frac{1}{2},k} \left( g _{x,i+\frac{1}{2},j+2,k} + g _{x,i-\frac{1}{2},j+2,k} - g _{x,i+\frac{1}{2},j+1,k} - g _{x,i-\frac{1}{2},j+1,k} \right. \nonumber \\
& & \left. - g _{x,i+\frac{1}{2},j,k} - g _{x,i-\frac{1}{2},j,k} + g _{x,i+\frac{1}{2},j-1,k} + g _{x,i-\frac{1}{2},j-1,k}\right) , 
\label{txyc} \\
4 \pi G \Delta T _{g,xy,i,j+\frac{1}{2},k} ^{(d)} & = & - \frac{\partial \phi}{\partial y} \frac{\partial ^3 \phi}{\partial x^3} \frac{h ^2}{6} \\
& = & - \frac{1}{24} g _{y,i,j+\frac{1}{2},k} \left( g _{x,i+\frac{3}{2},j+1,k} + g _{x,i+\frac{3}{2},j,k} - g _{x,i+\frac{1}{2},j+1,k} - g _{x,i+\frac{1}{2},j,k} \right. \nonumber \\
&& \left. - g _{x,i-\frac{1}{2},j+1,k} - g _{x,i-\frac{1}{2},j,k} + g _{x,i-\frac{3}{2},j+1,k} + g _{x,i-\frac{3}{2},j,k} \right) , \label{tgxxd}\\
4 \pi G \Delta T _{g,xy,i,j+\frac{1}{2},k} ^{(e)} & = & 
\frac{h^2}{12} \frac{\partial^2 \phi}{\partial x \partial z} \frac{\partial ^2 \phi}{\partial y \partial z}  \nonumber \\
& = & \frac{1}{96} \left( g _{x,i+\frac{1}{2},j+1,k+1} + g _{x,i+\frac{1}{2},j,k+1} + g _{x,i-\frac{1}{2},j+1,k+1} + g _{x,i-\frac{1}{2},j,k+1} - g _{x,i+\frac{1}{2},j+1,k} 
\right. \nonumber \\
&& \left. - g _{x,i+\frac{1}{2},j,k} - g _{x,i-\frac{1}{2},j+1,k} - g _{x,i-\frac{1}{2},j,k}
\right) 
\left( g _{y,i,j+\frac{1}{2},k+1} - g _{y,i,j+\frac{1}{2},k} \right) \nonumber \\
& & + \frac{1}{96} \left( g _{x,i+\frac{1}{2},j+1,k} + g _{x,i+\frac{1}{2},j,k} + g _{x,i-\frac{1}{2},j+1,k} + g _{x,i-\frac{1}{2},j,k} - g _{x,i+\frac{1}{2},j+1,k-1} 
\right. \nonumber \\
&& \left. - g _{x,i+\frac{1}{2},j,k-1} - g _{x,i-\frac{1}{2},j+1,k-1} - g _{x,i-\frac{1}{2},j,k-1}
\right) 
\left( g _{y,i,j+\frac{1}{2},k} - g _{y,i,j+\frac{1}{2},k-1} \right) . \label{Txy4e}
\end{eqnarray}
Equations (\ref{Txx4s}) through (\ref{Txx4e}) denote the diagonal component of the gravitational stress tensor on the cell surface centered at $\left( x _{i+\frac{1}{2}}, y _j, z _k \right) $, while Equations (\ref{Txy4s}) through (\ref{Txy4e}) describe a component of the gravitational stress tensor on the cell surface centered at $ \left( x _i, y _{j+\frac{1}{2}}, z _k  \right) $.  We obtain other components of the gravitational stress tensor by cyclic exchange of $ x $, $ y $, $ z $ and indices $ i $, $ j $, $ k $.

\section{Elimination of Spurious Circulation}
\setcounter{equation}{0}

The fourth order correction terms in the refined gravitational stress tensor are not unique (many such discretizations exist that are still of fourth order accuracy; c.f., HM24). In this Appendix, we justify our particular choice of the correction terms shown in Appendix A. 

The correction terms are either the quadrature of the second derivative of the gravitational potential or the product of the derivative and the third derivative, thus, the gravitational acceleration derived from the correction terms comprises products of the derivative and higher order derivatives, a part of which could originate from $ (\mathbf{\nabla} \times \mathbf{g})/(4 \pi G) \times \mathbf{g}$ (see Equation \ref{eq::divtg_exp}).  But such spurious terms should cancel with each other if the correction terms are well designed. 

As an example we examine $ \Delta T _{g,xx,i+\frac{1}{2},j,k} ^{(d)} $ and $ \Delta T _{g,xy,i,j+\frac{1}{2},k} ^{(c)} $. Both contain terms proportional to $ (\partial \phi /\partial y) $ (i.e., see Equations \ref{tgxxd} and \ref{txyc}). In the following we assume $ \partial /\partial z = 0 $ to simplify our inspection.
From Equation (\ref{tgxxa}) we obtain
\begin{eqnarray}
4 \pi G \left[ \Delta T _{g,xx,i+\frac{1}{2},j,k}^{(d)} - \Delta T _{g,xx,i-\frac{1}{2},j,k} ^{(d)}\right] & = &
\frac{1}{24} \left( g _{y,i+1,j+\frac{1}{2},k} - g _{y,i-1,j+\frac{1}{2},k} \right) \left( g _{y,i,j+\frac{3}{2},k} - 2 g _{y,i,j+\frac{1}{2},k} + g _{y,i,j-\frac{1}{2},k} \right) \nonumber \\
& + & \frac{1}{24} \left( g _{y,i+1,j-\frac{1}{2},k} - g _{y,i-1,j+\frac{1}{2},k} \right) \left( g _{y,i,j+\frac{1}{2},k} - 2 g _{y,i,j-\frac{1}{2},k} + g _{y,i,j-\frac{3}{2},k} \right) \nonumber \\
& + & \frac{1}{24} g _{y,i,j+\frac{1}{2},k} \left( g _{y,i+1,j+\frac{3}{2},k} - 2 g _{y,i+1,j+\frac{1}{2},k} + g _{y,i+1,j-\frac{1}{2},k} - g _{y,i-1,j+\frac{3}{2},k} \right. \nonumber \\
& & \left. + 2 g _{y,i-1,j+\frac{1}{2},k} - g  _{y,i-1,j-\frac{1}{2},k} \right) \nonumber \\
& + & \frac{1}{24} g _{y,i,j-\frac{1}{2},k} \left( g _{y,i+1,j+\frac{1}{2},k} - 2 g _{y,i+1,j-\frac{1}{2},k} + g _{y,i+1,j-\frac{3}{2},k} - g _{y,i-1,j+\frac{1}{2},k} \right. \nonumber \\
& & \left. + 2 g _{y,i-1,j-\frac{1}{2},k} - g  _{y,i-1,j-\frac{3}{2},k} \right) .
\label{tgxxe-x}
\end{eqnarray}
For later convenience we rewrite Equation (\ref{txyc}) as
\begin{eqnarray}
4 \pi G \Delta T _{g,xy,i,j+\frac{1}{2},k} ^{(c)}
& = & \frac{1}{24} g _{y,i,j+\frac{1}{2},k} \left( - g _{x,i+\frac{1}{2},j+2,k} - g _{x,i-\frac{1}{2},j+2,k} + 3 g _{x,i+\frac{1}{2},j+1,k} + 3 g _{x,i-\frac{1}{2},j+1,k} \right. \nonumber \\
& & \left. - 3 g _{x,i+\frac{1}{2},j,k} - 3 g _{x,i-\frac{1}{2},j,k} - g _{x,i+\frac{1}{2},j-1,k} - g _{x,i-\frac{1}{2},j-1,k}\right) \nonumber \\
&  & - \frac{1}{12} g _{y,i,j+\frac{1}{2},k}\left(  g _{x,i+\frac{1}{2},j+1,k} + g _{x,i-\frac{1}{2},j+1,k} - 2 g _{x,i+\frac{1}{2},j,k} - 2 g _{x,i-\frac{1}{2},j,k} + g _{x,i+\frac{1}{2},j-1,k} \right. \nonumber \\
& & \left. +g _{x,i-\frac{1}{2},j-1,k} \right) ,  \label{txyc-u} \\
4 \pi G \Delta T _{g,xy,i,j-\frac{1}{2},k} ^{(c)}
& = & - \frac{1}{24} g _{y,i,j-\frac{1}{2},k} \left( - g _{x,i+\frac{1}{2},j+1,k} - g _{x,i-\frac{1}{2},j+1,k} + 3 g _{x,i+\frac{1}{2},j,k} + 3 g _{x,i-\frac{1}{2},j,k} \right. \nonumber \\
& & \left. - 3 g _{x,i+\frac{1}{2},j-1,k} - 3 g _{x,i-\frac{1}{2},j-1,k} - g _{x,i+\frac{1}{2},j-2,k} - g _{x,i-\frac{1}{2},j-2,k}\right) \nonumber \\
&  & + \frac{1}{12} g _{y,i,j-\frac{1}{2},k}\left(  g _{x,i+\frac{1}{2},j+1,k} + g _{x,i-\frac{1}{2},j+1,k} - 2 g _{x,i+\frac{1}{2},j,k} - 2 g _{x,i-\frac{1}{2},j,k} + g _{x,i+\frac{1}{2},j-1,k} \right. \nonumber \\
& & \left. +g _{x,i-\frac{1}{2},j-1,k} \right) .  \label{txyc-l}
\end{eqnarray}
Combining Equations (\ref{txyc-u}) and (\ref{txyc-l}), we obtain
\begin{eqnarray}
4 \pi G \left[ \Delta T _{g,xy,i,j+\frac{1}{2},k} ^{(c)} -  \Delta T _{g,xy,i,j-\frac{1}{2},k} ^{(c)} \right] & = &
\frac{1}{24} g _{y,i,j+\frac{1}{2},k} \left( - g _{x,i+\frac{1}{2},j+2,k} - g _{x,i-\frac{1}{2},j+2,k} + 3 g _{x,i+\frac{1}{2},j+1,k}  \right. \nonumber \\
& & \left. + 3 g _{x,i-\frac{1}{2},j+1,k} - 3 g _{x,i+\frac{1}{2},j,k} - 3 g _{x,i-\frac{1}{2},j,k} - g _{x,i+\frac{1}{2},j-1,k} - g _{x,i-\frac{1}{2},j-1,k}\right) \nonumber \\
&& + \frac{1}{24} g _{y,i,j-\frac{1}{2},k} \left( - g _{x,i+\frac{1}{2},j+1,k} - g _{x,i-\frac{1}{2},j+1,k} + 3 g _{x,i+\frac{1}{2},j,k} + 3 g _{x,i-\frac{1}{2},j,k} \right. \nonumber \\
& & \left. - 3 g _{x,i+\frac{1}{2},j-1,k} - 3 g _{x,i-\frac{1}{2},j-1,k} - g _{x,i+\frac{1}{2},j-2,k} - g _{x,i-\frac{1}{2},j-2,k}\right) \nonumber \\
&& - \frac{1}{12} \left( g _{y,i,j+\frac{1}{2},k} - g _{y,i,j-\frac{1}{2},k} \right)\left(  g _{x,i+\frac{1}{2},j+1,k} + g _{x,i-\frac{1}{2},j+1,k} \right. \nonumber \\
&& \left. - 2 g _{x,i+\frac{1}{2},j,k} - 2 g _{x,i-\frac{1}{2},j,k} + g _{x,i+\frac{1}{2},j-1,k} +g _{x,i-\frac{1}{2},j-1,k} \right) . 
\label{tgxyc-y}
\end{eqnarray}
The first term in the right hand side of Equation (\ref{tgxyc-y}) cancels with the third term in the right hand side of Equation (\ref{tgxxe-x}) since
\begin{eqnarray}
     g _{y,i+1,j+\frac{3}{2},k} &  -  & 2 g _{y,i+1,j+\frac{1}{2},k} + g _{y,i+1,j-\frac{1}{2},k} - g _{y,i-1,j+\frac{3}{2},k} + 2 g _{y,i-1,j+\frac{1}{2},k} - g  _{y,i-1,j-\frac{1}{2},k} \nonumber \\
      & = & g _{x,i+\frac{1}{2},j+2,k} +  g _{x,i-\frac{1}{2},j+2,k} -  3 g _{x,i+\frac{1}{2},j+1,k} - 3 g _{x,i-\frac{1}{2},j+1,k} + 3 g _{x,i+\frac{1}{2},j,k} + 3 g _{x,i-\frac{1}{2},j,k} \nonumber \\
      & &  - g _{x,i+\frac{1}{2},j-1,k} - g _{x,i-\frac{1}{2},j-1,k} . \label{g-circit}
\end{eqnarray}
Both the left and right hand sides of Equation (\ref{g-circit}) are proportional to $ \partial ^4 \phi /[\partial x \partial y ^3]$.  The proportionality is demonstrated via terms containing $ g  _y $ (left hand side) and terms containing $ g _x $ (right hand side). We can confirm Equation (\ref{g-circit}) by substituting Equations (\ref{gx}) and (\ref{gy}).  Similarly the second term in the right hand side of Equation (\ref{tgxyc-y}) cancels with the fourth term in the right hand side of Equation (\ref{tgxxe-x}).  

Similar cancellation takes place when the $x$-derivative of $ \Delta T _{g,xx,i+\frac{1}{2},j,k} $
and the $y$-derivative of $ \Delta T _{g,xy,i,j+\frac{1}{2},k} $ are of the same form.  Often, these cancellations are associated with higher derivatives of $ \mathbf{\nabla} \times \mathbf{g} $.  Equation (\ref{g-circit}) denotes $ \partial ^2 (\mathbf{\nabla} \times \mathbf{g})/\partial y ^2 = 0. $  

In summary, considering circuits where $\oint \mathbf{g} \cdot d\mathbf{s} = 0$ provides an (admittedly complicated) avenue to making selections on particular discretizations that are (a) informed by the constraint $\mathbf{\nabla} \times \mathbf{g} = 0$ and (b) often simplify the resultant gravitaitonal source term due to nice cancellations.  In our derivations, we have taken painstaking measures to be consistent in this pattern, however, we cannot confidently claim that Equation (\ref{rhogx4b}) is completely devoid of rotation, however, if there is any such spurious circulation, it would only be in fourth order small quantities, and it certainly did not present itself in our tests (see, e.g., \S \ref{s+e}).  

\section{Conservation of Total Linear Momentum}
\label{ap:conservation}
\setcounter{equation}{0}

Upon substitution of $\rho_{i,j,k} \rightarrow \rho_{i,j,k}^{(2)}$, Equation (\ref{rhogx4b}) can be directly related to the divergence of a fourth-order accurate gravitational stress tensor. Without said substitution, Equation (\ref{rhogx4b}) cannot demonstrate such equivalence.  We here prove, however, that Equation (\ref{rhogx4b}) itself is still capable of conserving total linear momentum to round-off error.  For later convenience, we recast Equation (\ref{rhogx4b}) as
\begin{eqnarray}
\left( \rho g _x \right) _{i,j,k} & = & \left( \rho g _x \right) _{i,j,k} ^{(c)}
+ \left( \rho g _x \right) _{i,j,k} ^{(x)} + \left( \rho g _x \right) _{i,j,k} ^{(y+)} + \left( \rho g _x \right) _{i,j,k} ^{(y-)} + \left( \rho g _x \right) _{i,j,k} ^{(z+)} + \left( \rho g _x \right) _{i,j,k} ^{(z-)} ,\\
\left( \rho g _x \right) _{i,j,k} ^{(c)} & = & \frac{\rho _{i,j,k}}{12} \left( - g _{x,i+\frac{3}{2},j,k} + 7 g _{x,i+\frac{1}{2},j,k}
+ 7 g _{x,i-\frac{1}{2},j,k} - g _{x,i-\frac{3}{2},j,k} \right) ,  \\
\left( \rho g _x \right) _{i,j,k} ^{(x)} & = & \frac{\left( \rho _{i+1,j,k} - \rho _{i-1,j,k}\right)}{24} \left( g _{x,i+\frac{1}{2},j,k} - g _{x,i-\frac{1}{2},j,k} \right) , \\
\left( \rho g _x \right) _{i,j,k} ^{(y+)} & = &  \frac{(\rho _{i,j+1,k} - \rho _{i,j,k})}{48} \left( g _{x,i+\frac{1}{2},j+1,k}  + g _{x,i-\frac{1}{2},j+1,k} - g _{x,i+\frac{1}{2},j,k} - g _{x,i-\frac{1}{2},j,k} \right) , \\
\left( \rho g _x \right) _{i,j,k} ^{(y-)} & = & \frac{(\rho _{i,j,k} - \rho _{i,j-1,k})}{48} \left( g _{x,i+\frac{1}{2},j,k}  + g _{x,i-\frac{1}{2},j,k} - g _{x,i+\frac{1}{2},j-1,k} - g _{x,i-\frac{1}{2},j-1,k} \right) , \\
\left( \rho g _x \right) _{i,j,k} ^{(z+)} & = & \frac{(\rho _{i,j,k+1} - \rho _{i,j,k})}{48} \left( g _{x,i+\frac{1}{2},j,k+1} + g _{x,i-\frac{1}{2},j,k+1} - g _{x,i+\frac{1}{2},j,k} - g _{x,i-\frac{1}{2},j,k} \right) ,\\
\left( \rho g _x \right) _{i,j,k} ^{(z-)} & = & \frac{(\rho _{i,j,k} - \rho _{i,j,k-1})}{48} \left( g _{x,i+\frac{1}{2},j,k} + g _{x,i-\frac{1}{2},j,k} - g _{x,i+\frac{1}{2},j,k-1} - g _{x,i-\frac{1}{2},j,k-1} \right) .
\label{rhogx4c}
\end{eqnarray}
The term $  \left( \rho g _x \right) _{i,j,k} ^{(c)}$ is the product of the cell-volume-averaged density and gravity.  As shown in Equation (\ref{dPoisson}), the density is equivalent to a linear expression of the gravitational potential. The cell-volume-averaged gravity is equally a linear expression of the gravitational potential,
\begin{eqnarray}
\frac{- g _{x,i+\frac{3}{2},j,k} + 7 g _{x,i+\frac{1}{2},j,k}+7 g _{x,i-\frac{1}{2},j,k} - g _{x,i-\frac{3}{2},j,k}}{12} & = & 
\frac{\phi _{i+2,j,k} - 8 \phi _{i+1,j,k} + 8 \phi _{i-1,j,k} - \phi _{i-2,j,k}}{12 h} \\
& = & \sum _{i^\prime} \alpha _{i^\prime} \phi _{i+i^\prime,j,k} .
\end{eqnarray}
The coefficient is anti-symmetric with respect to $ i ^\prime $, i.e., $ \alpha _{-i^\prime} = - \alpha _{i^\prime} $.  Similarly we can express the density in the form of 
\begin{eqnarray}
\rho _{i,j,k} & = & \sum _{i^\prime} \sum _{j^\prime} \sum _{k ^\prime} \beta _{i^\prime,j^\prime,k ^\prime} \phi _{i+i^\prime,j+j^\prime,k+k^\prime} . 
\end{eqnarray}
The coefficient is symmetric with respect to $ i ^\prime $, $ j ^\prime $, and $  k ^\prime $, i.e., $ \beta _{\pm i^\prime,\pm j ^\prime, \pm k ^\prime} = \beta _{i^\prime,j^\prime, k ^\prime}$. Because of this symmetry, the sum of $ \left( \rho g  _x \right) _{i,j,k} ^{(c)} $ vanishes.  Similarly, $ \left( \rho g  _x \right) _{i,j,k} ^{(x)} $ is the product of two linear expressions of the gravitational potential.  One is symmetric with respect to the cell center $(i,j,k)$, while the other is anti-symmetric in the $ x $-direction. Thus, the sum of $ \left( \rho g _x \right) _{i,j,k} ^{(x)} $ also vanishes. The term $ \left( \rho g _x \right) _{i,j,k} ^{(y+)} $ is also the product of two linear expression of the gravitational potential.  Both of them are either symmetric or anti-symmetric about the cell surface center $(i,j+\frac{1}{2},k) $. Thus, again the sum vanishes.  The other terms also vanish according to the similar symmetries.  Since the sum of the gravitational acceleration vanishes, the total linear momentum is conserved to round-off error.

\bibliography{selfG}{}
\bibliographystyle{aasjournal}

\end{document}